\documentclass{aa}  
\usepackage{graphicx}
\usepackage{natbib}
\usepackage{color}
\usepackage{upgreek}
\usepackage{soul}

\bibpunct{(}{)}{;}{a}{}{,}
\newcommand{\teff}{T_{\rm eff}}
\newcommand{\kms}{km\,s$^{-1}$}

\newcommand{\vsini}{v \sin i}
\newcommand{\bz}{\langle B_{\rm z} \rangle}

\usepackage{txfonts}
\usepackage{mathrsfs}
\begin{document}

\titlerunning{Magnetosphere of HD\,190073}
\authorrunning{J\"arvinen et al.}

\title{
First snapshot of a magnetosphere around a Herbig~Ae/Be star
}

   \author{
          S.~P.~J\"arvinen\inst{1}
          \and
          S.~Hubrig\inst{1}
          \and
          M.~K\"uker\inst{1}
          \and
          U.~Ziegler\inst{1}
          \and
          I.~Ilyin\inst{1}
          \and
          M.~Sch\"oller\inst{2}
          \and
          H.~Adigozalzade\inst{3}
          \and
          N.~Ismailov\inst{3}
          \and
          U.~Bashirova\inst{3}
          }

   \institute{Leibniz-Institut f\"ur Astrophysik Potsdam (AIP),
              An der Sternwarte~16, 14482~Potsdam, Germany\\
              \email{sjarvinen@aip.de}
            \and 
          European Southern Observatory, 
              Karl-Schwarzschild-Str.~2, 85748~Garching, Germany
         \and
             N.\ Tusi Shamakhy Astrophysical Observatory, 
             settl.\ Y.\ Mammadaliyev, Shamakhy, Azerbaijan, AZ 5626
}

   \date{Received 11 14, 2024; accepted 03 13, 2025}

 
  \abstract
   {
The Herbig Ae/Be star HD\,190073 is one of the very few magnetic Herbig 
Ae/Be stars for which close low-mass companions have been reported. Previously 
published magnetic field measurements indicated an annual change in the 
field configuration.
}
{
We aim to study in detail the spectral and magnetic variability of this star 
and characterise its magnetosphere for the first time. 
}
{
Newly acquired and archival spectropolarimetric observations are combined to 
determine a more precise magnetic period and to constrain the geometry of 
the magnetic field. The variability of hydrogen line profiles is studied 
using dynamical spectra. Archival X-shooter observations of the 
\ion{He}{i}\,10\,830\,\AA{} triplet are used to characterise its variability 
over the rotation cycle. Further, we carry out 2D magnetohydrodynamical 
(MHD) simulations of the magnetosphere using the {\sc Nirvana} MHD code. 
}
{
From the spectropolarimetric observations, we determine for HD\,190073 a 
magnetic period $P=51.70\pm0.06$\,d. We estimate a magnetic obliquity 
angle $\beta=82.9\pm6.4^{\circ}$ and a dipole strength 
$B_{\rm d}=222\pm66$\,G. Our dynamical spectra constructed for the hydrogen 
line profiles observed during 2011 clearly reveal a ringlike magnetospheric 
structure appearing at the rotation phase of best visibility of the positive 
magnetic pole. These spectra present the first snapshot of a magnetosphere 
around a Herbig Ae/Be star. 2D MHD simulations involving nonisothermal gas 
show that the magnetosphere is compact, with a radius of about $3\,R_*$, and 
that the wind flow extends over tens of $R_*$. With a reported radius of the 
accretion disk of 1.14\,au around HD\,190073, the distance between the star 
and the disk is about 25\,$R_*$. The detection of a magnetosphere around 
HD\,190073, and the possible presence of lower-mass companions at different 
distances, make this system a valuable laboratory for studying the magnetic 
interaction between the host star, its companions, and the accretion disk.
   }
   {}


   \maketitle
%

\section{Introduction}\label{sec:introduction}

Herbig Ae/Be stars are pre-main-sequence objects with pronounced emission 
line features and an infrared excess indicative of dust in their 
circumstellar disks. There is observational evidence that Herbig Ae and 
late-type Herbig Be stars are intermediate-mass analogues of T\,Tauri stars, 
but with convectively stable envelopes that do not support the dynamo 
action found in the fully convective T\,Tauri stars 
\citep[e.g.][]{HubrigSchoeller2021}.
A number of magnetic studies have been attempted in the last few years, 
indicating that about two dozen Herbig Ae/Be stars very probably possess 
globally organised magnetic fields reminiscent of fields of the classical 
Ap/Bp stars, of which the magnetic Herbig stars may be the precursors
\citep[e.g.][]{Hubrig2004,Alecian2013a,Hubrig2015,Silva2015,Silva2018,Silva2019}.

Magnetic fields in these stars might be fossils of the early star formation
epoch, in which the magnetic field of the parental magnetised core was  
compressed into the innermost regions of the accretion disks
\citep[e.g.][]{BanPud2006}.
Alternatively, it is plausible that the weak magnetic fields detected in a 
number of Herbig Ae/Be stars are just leftovers of the fields generated by 
pre-main-sequence dynamos during the convective phase
\citep[e.g.][and references therein]{HubrigSchoeller2021}.
If this scenario is 
valid, we should expect a significantly larger number of Herbig stars 
possessing weak magnetic fields. Indeed, fields greater than 200\,G are very 
rare, with most stars possessing fields of about 100\,G and less 
\citep{Hubrig2015}. 
The studies by 
\citet{Silva2018,Silva2019}
also support this scenario: the authors report the presence of relatively 
weak longitudinal magnetic field strengths in a few Herbig Ae/Be stars in 
the range from 120\,G to 210\,G.

To understand the origin of magnetic fields in Herbig Ae/Be stars, knowledge 
of the field geometry is indispensable, but requires information on the 
stellar rotation period. Previous studies of upper main-sequence stars with
large-scale organised magnetic fields demonstrated that their magnetic 
fields are dominated by dipolar fields tilted with respect to the
rotation axis
\citep[following the oblique dipole rotator model;][]{Stibbs1950}.
Thus, their rotation periods  were frequently determined by studying the 
periodicity in the available magnetic data.

Currently, the magnetic periods and corresponding magnetic phase curves 
presenting the dependence of mean longitudinal magnetic field strength on 
the rotation phase have only been reported for three Herbig 
Ae/Be stars, 
\object{HD\,101412} 
\citep{Hubrig2011a},
\object{V380\,Ori} 
\citep{Alecian2009}, 
and 
\object{HD\,190073} 
\citep[=V1295\,Aql;][]{Alecian2013b}.
However, V380\,Ori has not been confirmed to be a Herbig star and is 
probably a very young main-sequence star
\citep{Hubrig2019, Shultz2021}.
A rotation period of 42.076\,d for HD\,101412 was reported by 
\citet{Hubrig2011},
who used a combination of photometric observations and measurements of the 
longitudinal magnetic field based on spectropolarimetric observations 
with the low-resolution FOcal Reducer low dispersion Spectrograph 
\citep[FORS\,1;][]{FORS1} 
at the Very Large Telescope (VLT). This star has also been reported to 
possess a strong surface magnetic field of 3.5\,kG 
\citep{Hubrig2010}. 
The measured mean longitudinal magnetic field exhibits a single-wave variation 
during the stellar rotation cycle. Such a variability is usually considered 
as evidence for a dominant dipolar contribution to the magnetic field 
geometry. Spectroscopic signatures of magnetospheric accretion in HD\,101412 
have been discussed by 
\citet{MAHD101412}.

The presence of a magnetic field on the surface of the Herbig Ae/Be star 
HD\,190073 has been known of for several years. The first measurement of a 
longitudinal magnetic field in HD\,190073 was published by 
\citet{Hubrig2006},
who indicate the presence of a longitudinal magnetic field 
$\bz = 84\pm30$\,G  at a 2.8$\sigma$ level, measured with FORS\,1. 
Subsequently, this star was studied by 
\citet{Catala2007}. 
The authors used observations with the Echelle SpectroPolarimetric Device 
for the Observation of Stars 
\citep[ESPaDOnS;][]{Donati2006proc} 
installed at the Canada-France-Hawaii Telescope, at a spectral resolution 
of $R\approx65\,000$ and, employing the least-squares deconvolution (LSD) 
introduced by 
\citet{Donati1997}, 
confirmed the presence of a weak longitudinal magnetic field, 
$\bz = 74\pm10$\,G, at a higher significance level. The applied LSD method 
allows us to construct average photospheric profiles of both the Stokes~$I$ 
and Stokes~$V$ parameters, and the diagnostic null $N$ profile, by 
deconvolving the observed spectra using a line mask with lines identified in 
these spectra.

Additional observations of HD\,190073 using the Narval spectropolarimeter 
\citep{Narval} 
at the 2\,m Telescope Bernard Lyot at Pic du Midi and the High Accuracy Radial 
velocity Planet Searcher polarimeter 
\citep[HARPS\-pol;][]{Harps}
at the 3.6\,m European Southern Observatory (ESO) telescope on La~Silla have 
subsequently been obtained by
\citet{Alecian2013b}
and
\citet{Silva2015,Silva2019} 
between 2011 and 2019. Using the numerous observations of this star during 
2011 May--November and 2012 July--October, and employing the LSD technique,
\citet{Alecian2013b}
revealed variations of the magnetic field strength on timescales of days to 
weeks. Furthermore, in contrast to the 2011 measurements, which showed a 
longitudinal magnetic field of positive polarity for all but one 
measurement, a sizeable number of measurements in 2012 indicated a field of 
negative polarity. To explain the observed annual change in the magnetic field 
configuration, the authors suggested that a phenomenon such as an 
interaction between the fossil field and the ignition of a dynamo field 
generated in the newly born convective core occurred, perturbing the 
magnetic field. In the study of 
\citet{Alecian2013b}, 
the best fit to the 2011--2012 data set corresponding to a rotation 
period of $39.8\pm0.5$\,d reproduced all but one of the 2012 observations, 
but a number of the 2011 data points were discordant with this period. No 
significant period could be identified in the 2011 data set, while a period 
of $40\pm5$\,d was clearly detected in the 2012 data set when analysing 
these data sets separately.

Since HD\,190073 is rather bright, with m$_{\rm V}=7.7$, it has been 
intensively studied in recent years in various wavelength domains. More 
accurate fundamental parameters have recently been reported by 
\citet{GuzmanDiaz2021} 
from the analysis of the spectral energy distribution and Gaia Early Data 
Release 3 parallax and photometry 
\citep{Riello2021}.
The following stellar parameters have been derived: spectral type B9, 
effective temperature $\teff=9750\pm125$\,K, mass $M=6.0\pm0.2\,M_{\sun}$, 
radius $R=9.68\pm0.44\,R_{\sun}$, and age $0.30\pm0.02$\,Myr. In their study,
\citet{Aarnio2017} used a radial velocity $v_{r}=-1.2\pm1.3$\,\kms{} and 
$\vsini=3.19\pm2.45$\,\kms.

The presence of a super-Jupiter-mass candidate around HD\,190073 at a 
separation of about 1.1\arcsec{} has been announced by 
\citet{Rich2022} 
using the Gemini Planet Imager 
\citep{GPI}.
More recently, 
\citet{Ibrahim} 
studied HD\,190073 using the Center for High Angular Resolution Astronomy 
(CHARA) and the Very Large Telescope Interferometer (VLTI) arrays. The 
authors report that their modelling is consistent with a near face-on disk 
with an inclination $\le$\,20$^{\circ}$ and an average radius of a ringlike 
structure around the star of $1.4\pm0.2$\,mas (1.14\,au), which is 
interpreted as the dust destruction front. The observations indicate the 
disk around HD\,190073 to have skewness on the finer spatial scale, possibly 
due to a low-mass companion. The sub-au structure that is detected in the 
image seems to move between the two epochs inconsistently, with Keplerian 
motion, pointing at dynamics effects from the outer disk. The authors 
speculate that the motion could be caused by interactions in the outer disk 
with an object embedded in the inner disk.

In addition, X-ray emission of $\log L_x = 29.86\pm0.34$\,erg\,s$^{-1}$ has 
recently been detected by 
\citet{Anilkumar} 
from observations with the {\em Chandra} X-ray Observatory 
\citep{Chandra}. 
The most popular mechanisms proposed to explain X-ray emission from Herbig 
Ae/Be stars involve contribution from a low-mass companion in the close 
proximity of the Herbig Ae/Be stars 
\citep[e.g.][]{Stelzer},
wind shocks, or magnetically confined winds 
\citep[e.g.][]{GandS2009}.
These very recent findings make a reinvestigation of the magnetic properties
of HD\,190073 especially invaluable in view of a possible magnetospheric 
interaction between the magnetic host star, its low-mass companion, and the 
accretion disk. As for the giant planetary companion, additional observations 
are needed to confirm that it is co-moving with the host star.

After obtaining additional HARPS\-pol observations, we decided to reexamine 
the magnetic field configuration of HD\,190073. This included a careful 
inspection of the available spectropolarimetric data and the determination 
of the rotation period based on our own magnetic field measurements. As the 
investigation of the rotation modulation of the hydrogen emission lines 
using dynamical spectra phased with a redetermined rotation period suggests 
the presence of a compact magnetosphere, we characterised this magnetosphere 
using the Nirvana MHD code developed in the Leibniz-Institut f\"ur 
Astrophysik Potsdam 
\citep{Nirvana1,Nirvana2}. 
Further, we studied the rotational modulation of a few metal lines and of the 
\ion{He}{i} 10\,830\,\AA{} triplet, usually used as a diagnostic of 
magnetically driven accretion and outflows.

This paper is structured as follows:
details about the observations of HD\,190073 considered in this study are 
described in Sect.~\ref{sec:obs}. In Sect.~\ref{sec:Bz}, we present our 
magnetic field measurements that we used to determine the rotation period of 
HD\,190073. In Sect.~\ref{sec:vari}, we study the line profile variability 
of several spectral lines over the rotation period, including the 
variability of hydrogen emission lines, which clearly shows the presence of a 
magnetospheric emitting region. The parameters of the detected magnetosphere 
are explored in Sect.~\ref{sec:MHD} and the results obtained are 
discussed in Sect.~\ref{sec:dis}.


\section{Observational material}\label{sec:obs}

\subsection{High-resolution HARPS\-pol observations}

Seven high-resolution ($R\approx115\,000$) polarimetric spectra of HD\,190073 
have been recorded with HARPS\-pol between 2011 and 2017 
\citep{Silva2015,Silva2019}. 
Two more observations were acquired with this instrument in 2019, on June 
16 and 18. The latter has been omitted from our analysis due to a very low
signal-to-noise ratio. Polarimetric HARPS\-pol spectra cover wavelengths 
from 3780 to 6910\,\AA{}, with a small gap between 5259 and 5337\,\AA. The 
observations obtained were reduced using a dedicated HARPS data reduction 
pipeline available at the ESO La Silla observatory. Details on the 
normalisation of the spectra to the continuum level can be found in 
\citet{Hubrig2013}.

\subsection{Archival observations}

The reduced Narval observations described in the work of 
\citet{Alecian2013b}
were downloaded from the PolarBase 
\citep{Polarbase}
archive\footnote{\url{http://polarbase.irap.omp.eu/}}.
Polarimetric spectra obtained with this instrument have a spectral resolution 
$R\approx65\,000$ and cover wavelengths from 3750 to 10500\,\AA. The data set 
of the available spectral material for 2011 contains 40 polarimetric 
spectra, whereas for 2012 the data set includes 25 spectra. The numbers of 
observations in 2011 and 2012 reported in the work of 
\citet{Alecian2013b} 
were 31 and 20, respectively. We also downloaded one available ESPaDOnS 
spectrum obtained in 2012. ESPaDOnS can be considered as a twin brother of 
Narval with the same wavelength coverage and a similar spectral resolution.

Furthermore, to investigate the rotational variability of the 
\ion{He}{i}\,10\,830\,\AA{} triplet with the transition 
$2s^3S_1-2p^3P^0_{0,1,2}$ we collected archival X-shooter observations 
of HD\,190073 acquired in the years 2010, 2015, and 2017. It has previously 
been demonstrated that in magnetic Herbig Ae/Be stars the line parameters of 
the \ion{He}{i}\,10\,830\,\AA{} triplet show cyclic variability related to 
the stellar magnetic period 
\citep[e.g.][]{MAHD101412}.
The spectral data with this instrument are obtained simultaneously over the 
entire spectral range from the near-UV to the near-IR in three different 
arms. The spectral resolution in the red arm is $R\approx11\,000$. X-shooter 
is operated by ESO on the VLT on Cerro Paranal, Chile. 


\section{Magnetic field measurements and the search for rotational modulation}
\label{sec:Bz}

As in our previous studies using HARPS\-pol data 
\citep[see e.g.][]{Silva2018,Silva2019},
to increase the accuracy of the mean longitudinal magnetic field ($\bz$) 
determination, we employed the singular value decomposition (SVD) technique,
following the description given by 
\citet{SVD}.
The mean longitudinal magnetic field is usually determined by computing 
the first-order moment of the SVD Stokes~$V$ profile according to
\citet{Mathys1989}.
The parameters of the lines used to calculate the SVD profiles were taken
from the Vienna Atomic Line Database 
\citep[VALD3;][]{Kupka2011}.
Only lines that appear to be unblended or minimally blended in the 
Stokes~$I$ spectra were included in the line mask. The resulting profiles 
were scaled according to line strength and sensitivity to the magnetic 
field. The final line mask contains 688 lines and includes Si, Sc, Ti, V, 
Cr, Mn, Fe, Ni, and Zr lines. The presence of a magnetic field in the SVD 
profile was evaluated according to 
\citet{Donati1997},
who defined that a Zeeman profile with a false alarm probability (FAP) 
$\leq 10^{-5}$ is considered as a definite detection, 
$10^{-5} <$ FAP $\leq 10^{-3}$ as a marginal detection, and 
FAP $> 10^{-3}$ as a nondetection.

An inspection of the 65 Narval SVD Stokes~$I$, $V$, and $N$ profiles 
acquired in 2011 and 2012 reveals that a few observations are either too 
noisy and therefore not reliable (2011 Oct 14, 2012 Jul 23, and 2012 Nov 20)
or show strong deviations in the profile shapes recorded on consecutive 
nights (observed in 2011 on Jul 4, 7, and 8 as well as on Oct 10, 11, and 
12). The noisy profiles were immediately excluded from the analysis, whereas
the deviating profiles were excluded at a later point.

\begin{figure}
\centering
\includegraphics[width=0.43\textwidth]{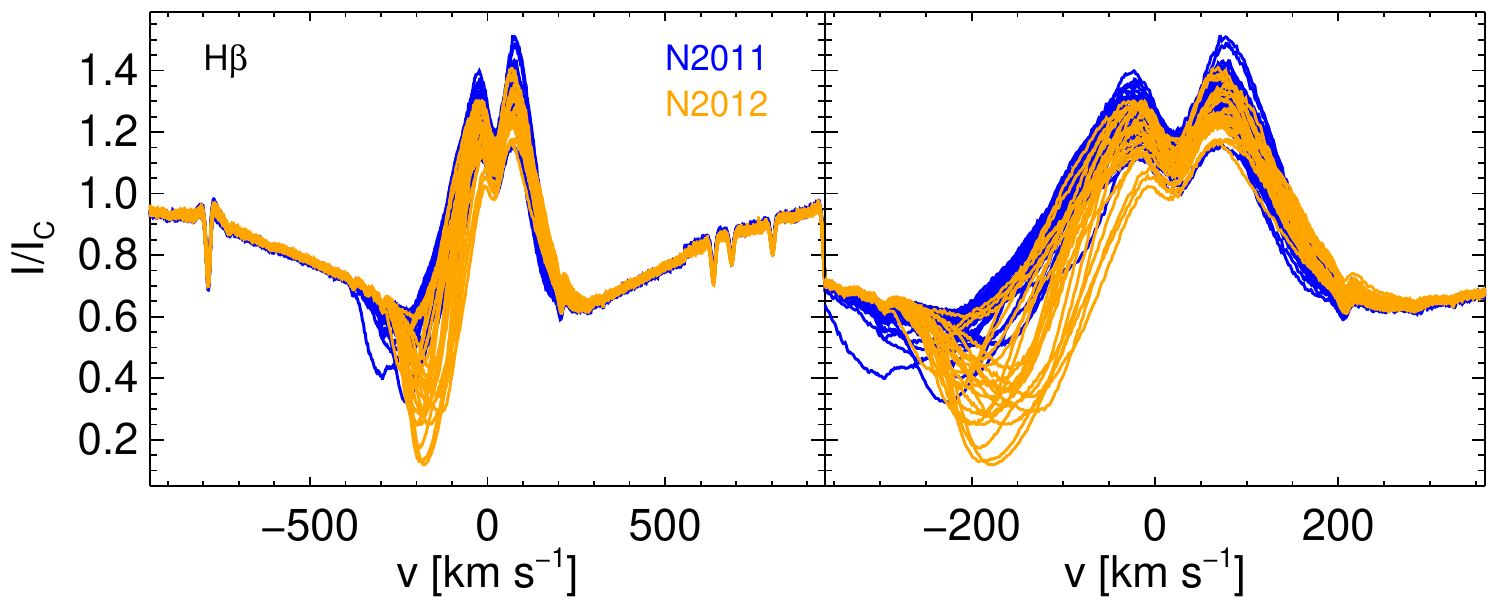}
\includegraphics[width=0.43\textwidth]{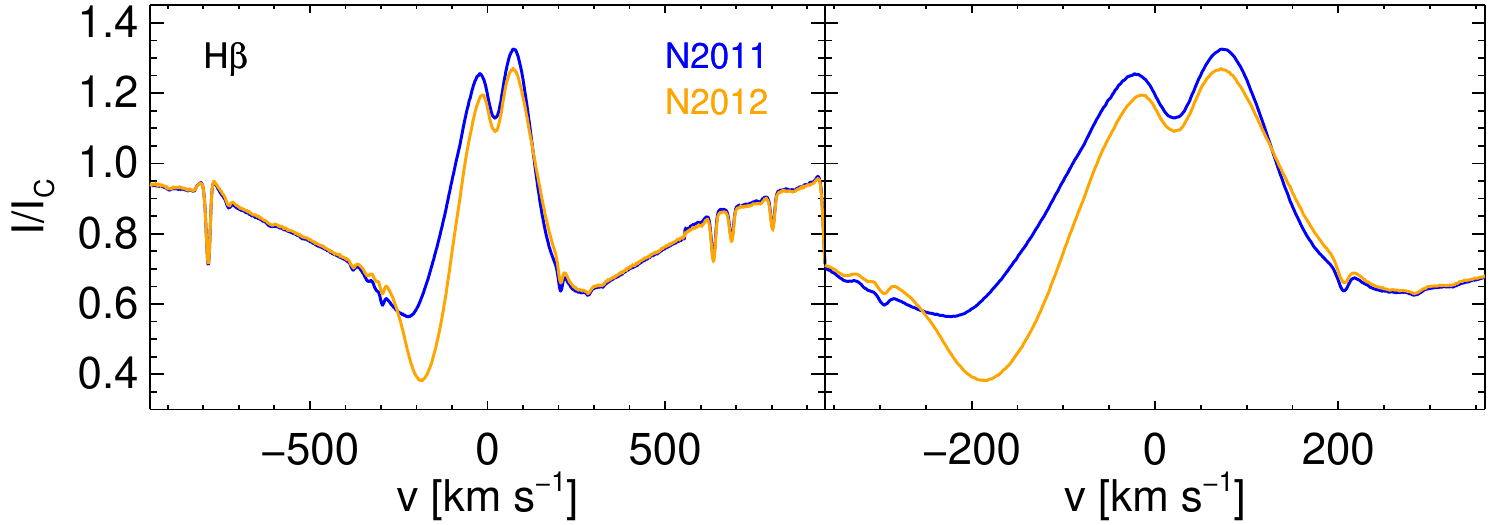}
\caption{
Variability of H$\beta$ profiles in Narval observations.
{\it Top panels:} Overplotted Narval H$\beta$ profiles from 2011 (blue) and 
2012 (yellow).
{\it Bottpm panels:} Mean  H$\beta$ profiles for both years. The
plots on the left panels include the full wings of the H$\beta$ line whereas
the plots on the right side show the central part of the profiles.
}
\label{fig:Hbeta1112}
\end{figure}

\begin{table}
\centering
\caption{
Different period analyses.
}
\label{tab:Bzper}
\begin{small}
\begin{tabular}{llcr@{$\pm$}lc}
\hline\hline\noalign{\smallskip}
Measurements     & N & f        & \multicolumn{2}{c}{$P$} & FAP                \\
                 &   & $[$d$^{-1}]$         & \multicolumn{2}{c}{$[$d$]$}   &                    \\
\hline\noalign{\smallskip}
\multicolumn{6}{c}{Period analysis of our $\bz{}$ measurements} \\
\hline\noalign{\smallskip}
2011 N           & 37 & 0.01773 & 56.4 & 2.7              & 1.2$\times 10^{-3}$ \\
2012 N           & 23 & 0.01938 & 51.6 & 1.1              & 9.5$\times 10^{-5}$ \\
2011 + 2012 N    & 60 & 0.02522 & 39.6 & 0.2              & 5.8$\times 10^{-6}$ \\
2012 N + H + E   & 34 & 0.01933 & 51.7 & 0.06             & 1.3$\times 10^{-8}$ \\
\hline\noalign{\smallskip}
\multicolumn{6}{c}{Period analysis from \ion{Fe}{ii} 6149\,\AA{} EWs} \\
\hline\noalign{\smallskip}
2012 N            & 23 & 0.01955 & 51.1 & 0.6 & $< 10^{-10}$\\
\hline
\end{tabular}
\end{small}
\end{table}

Already, a cursory review of the shapes of the hydrogen line profiles in 
the Stokes~$I$ spectra reveals striking differences between the profiles 
recorded in 2011 and in 2012, probably pointing to a long-term variability 
caused by an interaction with an unknown component in the HD\,190073 system. 
In Fig.~\ref{fig:Hbeta1112} we present the observed overplotted and mean 
H$\beta$ profiles. The absorption component of the P\,Cygni profiles of 
H$\beta$, related to wind outflow and shifted to the blue by about 
200\,\kms{}, is significantly deeper and the emission component of the 
profiles is lower in 2012 compared to the profile shapes recorded in 2011.
Since HD\,190073 possesses a globally organised magnetic field, we assume 
that the emission component of the profile is related to a magnetosphere 
around HD\,190073. Following the results of the abovementioned studies 
of this star, we speculate that the observed annual change in the hydrogen 
line profiles from 2011 to 2012 is due to an interaction with a secondary 
body in the system. Such an interaction could cause an increase in the wind 
outflow, producing partial screening of the magnetosphere. With respect to 
the marked variability of the hydrogen line profiles on a daily timescale,
we discuss in Sect.~\ref{sec:vari} that this variability can be
caused by a change of the viewing angle between the observer and the 
magnetosphere during stellar rotation. Similar changes in hydrogen line 
profiles over a rotation period are frequently observed in studies of 
magnetic O and B stars and usually used to determine magnetic periods 
\citep[e.g.][]{Kueker2024}.

Our period search was performed using a nonlinear least-squares fit 
to the multiple harmonics utilising the Levenberg–Marquardt method  
\citep{press}. 
To detect the most probable period, we calculated the frequency spectrum 
with a specific number of trial frequencies within the region of interest. A 
weighted linear least-squares fit was used for each frequency to fit a sine 
curve and bias offset. A sine wave is expected if the magnetic field is 
similar to a dipole field inclined with respect to the rotation axis. Based on 
the result of the fit, we performed a statistical test to check the null 
hypothesis on the absence of periodicity, that is, to check the statistical 
significance of the amplitude of the fit 
\citep{seber}.
The resulting F-statistics can be thought of as the total sum, including
covariances of the ratio of the harmonic amplitudes to their standard 
deviations. 

\begin{figure*}
\centering
\includegraphics[width=0.33\textwidth]{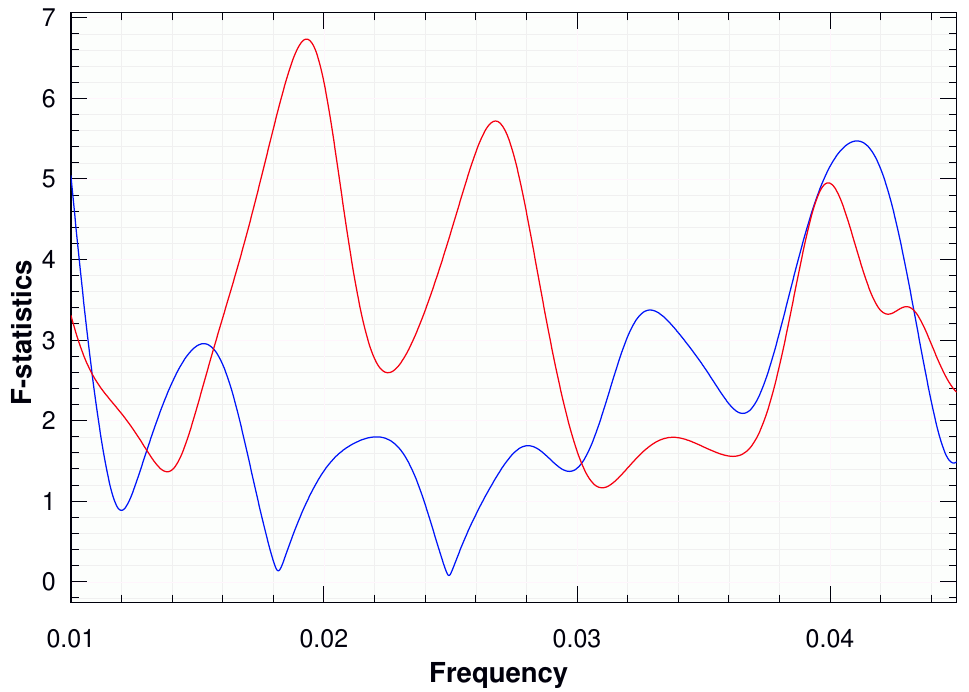}
\includegraphics[width=0.33\textwidth]{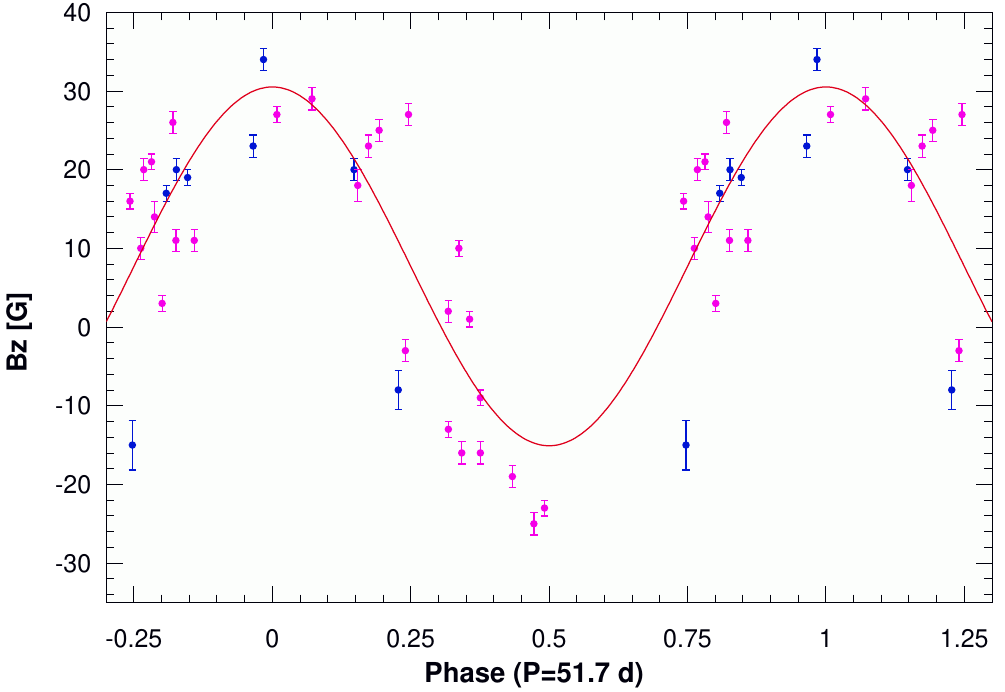}
\caption{
Periodogram and the magnetic phase curve for the period $P=51.7\pm0.06$\,d.
{\it Left:} Periodogram based on all Narval and one ESPaDOnS $\bz{}$
measurements from 2012 and eight HARPS\-pol $\bz{}$ measurements recorded
until 2019. The red line is for observations and the blue line for the
window function.
{\it Right:} $\bz{}$ measurements (magenta circles for Narval and ESPaDOnS
and blue circles for HARPS) phased with $P=51.7\pm0.06$\,d. The sinusoidal
fit is shown with the red line.
}
\label{fig:Bz2012H}
\end{figure*}

Using only the $\bz{}$ measurements obtained from Narval spectra during the 
years 2011 and 2012, we obtain a period $P=39.6$\,d 
(FAP $5.8\times 10^{-6}$), which is in agreement with the period reported by
\citet{Alecian2013b}.  
The results of our period search in these data are presented in 
Fig.~\ref{afig:Bz20112012}. Similar to the study of 
\citet{Alecian2013b}, 
we do not find any significant period in the data acquired in 2011 
(Fig.~\ref{afig:Bz2011}). For the Narval observations recorded in 2012, we 
obtain a definite detection of periodicity $P=51.6$\,d (FAP 
$9.5\times 10^{-5}$; Fig.~\ref{afig:Bz2012}). Including in this data set 
additional $\bz{}$ measurements using HARPS\-pol and ESPaDOnS observations, 
an even better FAP ($1.3\times 10^{-8})$ refers to the period $P=51.7$\,d. 
The corresponding periodogram based on these observations and the phase 
curve are presented in Fig.~\ref{fig:Bz2012H}. The summary of our search 
for periodicities in our $\bz{}$ measurements using different data sets is 
presented in Table~\ref{tab:Bzper}, which also includes the period search based
on metal lines described in Sect.~\ref{sec:metal}. The ephemeris based on 
the best visibility of the positive magnetic field extremum is described as
\begin{equation}
        T_{\mathrm{max(BsMJD)}} = 56108.06 + 51.70(6)\times E.
\end{equation}

A minimum dipole strength, $B_{\rm d}$, of 102\,G is estimated for HD\,190073 
employing the relation $B_{\rm d} \ge 3 \left| \bz_{\rm ,all} \right|$ 
\citep{Babcock1958},
with a strongest measured $\bz{}$ of $34\pm 3$\,G. Assuming a rotation 
period, $P$, of $51.7 \pm 0.06$\,d and a stellar radius, $R_\ast$, of 
$9.68 \pm 0.44\,R_\odot$ from 
\citet{GuzmanDiaz2021}, 
we obtain an equatorial velocity $v_{\rm eq} = 9.47 \pm 0.43$\,km\,s$^{-1}$.
Using $\vsini = 3.19 \pm 2.45$\,\kms{} from the study of 
\citet{Aarnio2017},
the inclination, $i$, that is, the angle between the rotation axis and the line 
of sight, is $19.7 \pm 15.8^{\circ}$, which is not very accurate, but fully 
in agreement with interferometric observations 
\citep{Ibrahim}.

The general description for the strength of the observed longitudinal 
magnetic field for a simple centred dipole was presented by 
\citet{Preston1967}:
\begin{equation}
\left< B_{\rm z} \right> = \frac{1}{20} \, \frac{15+u}{3-u} \, B_d (\cos \beta \cos i + \sin \beta \sin i \cos 2\pi t / P)
\label{eq:diagn.oblique_mod},
\end{equation}
where $\bz{}$ is the longitudinal magnetic field,
$\beta$ is the angle between the rotation axis and the magnetic axis,
$i$ is the angle between the rotation axis and the line of sight,
$P$ is the rotation period of the star, $u$ is the limb-darkening coefficient, and 
$B_d$ the strength of the dipolar magnetic field.
The relative amplitude of variation of the fitted longitudinal magnetic 
field phase curve is usually characterised by the parameter $r$, representing 
the ratio between $\bz_{\rm min}$ and $\bz_{\rm max}$. In our case, with 
$\bz_{\rm max}=31\pm3$\,G and $\bz_{\rm min}=-15\pm3$\,G, we have 
$r=-0.48\pm0.11$. Using
\begin{equation}
r  =  \frac{\bz_{\rm min}}{\bz_{\rm max}}
=  \frac{\cos \beta \cos i - \sin \beta \sin i}{\cos \beta \cos i + \sin \beta \sin i}
\label{eq:diagn.oblique_r},
\end{equation}
we calculated an obliquity angle $\beta=82.9\pm6.4^{\circ}$. Assuming a 
limb-darkening coefficient of 0.5 
\citep{Claret2019}, 
we obtain a polar magnetic field strength of $B_{\rm d}=222\pm66$\,G, 
typical for Herbig Ae/Be stars 
\citep[e.g.][]{Hubrig2015}.

\section{Line profile variability}\label{sec:vari}

The strong variability of the hydrogen emission lines on timescales of days to 
years in the spectra of HD\,190073 has already been reported in a number of 
studies 
\citep[e.g.][]{Aarnio2017,Kozlova2019,Hemayil}, 
but the reason for such a behaviour has remained unclear. Rotational modulation 
of a magnetically confined wind as the most likely origin for the 
variability of emission Balmer lines observed in stars possessing magnetic 
fields was suggested by 
\citet{Donati2006}.
Later on, 
\citet{Sundqvist2012} 
calculated the H$\alpha$ line profile for more than 100 snapshots of a 2D 
MHD simulation for a magnetic massive star with a dynamical magnetosphere 
and demonstrated a good overall agreement between their model and 
observations. The presented simulations strongly supported the assumption 
that the dynamical magnetospheric model captures the key physics responsible 
for the H$\alpha$ variability.

To try to understand the variability character of the hydrogen lines of 
HD\,190073, in addition to the Narval and ESPaDOnS observations from 
2011--2012, we downloaded from the PolarBase archive all available Narval 
and ESPaDOnS observations distributed over the years 2005--2008, 2013, and 
2016-2017. We analysed them together with the HARPS\-pol spectra acquired 
between 2011 and 2019. Unfortunately, since only single or very few spectra 
have been recorded per year, it is impossible to get a complete picture of the 
variability on the long timescale necessary to understand whether an 
interaction with a second body in the system indeed takes place and at which 
time intervals. Our discussion of the observed most extreme changes in the 
line profiles in the period from 2005 to 2019, and the corresponding figures, 
are presented in Fig.~\ref{afig:Hbetayr}. In addition, we 
show the variability observed in the line profiles based on medium-resolution 
spectra, which are described and discussed in Appendix~\ref{bsec:vari} and 
are displayed in Fig.~\ref{afig:Hbetamed}.

\begin{figure*}
\centering
\includegraphics[width=0.26\textwidth]{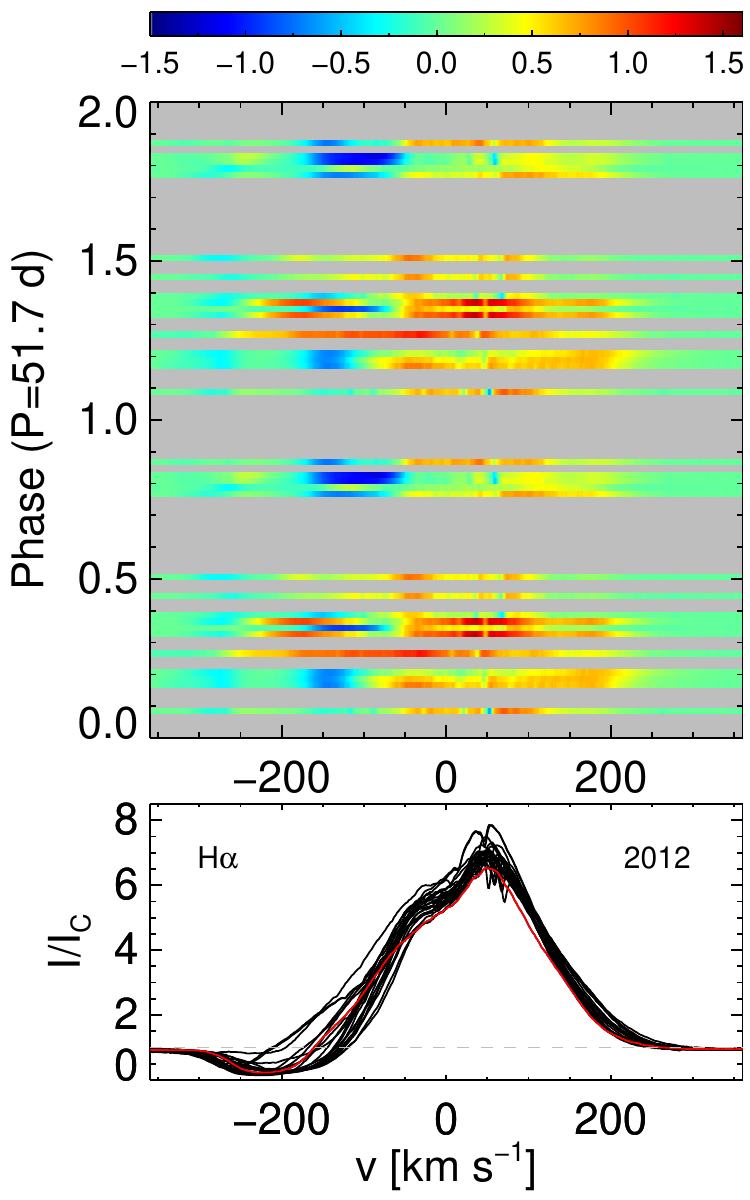}
\includegraphics[width=0.26\textwidth]{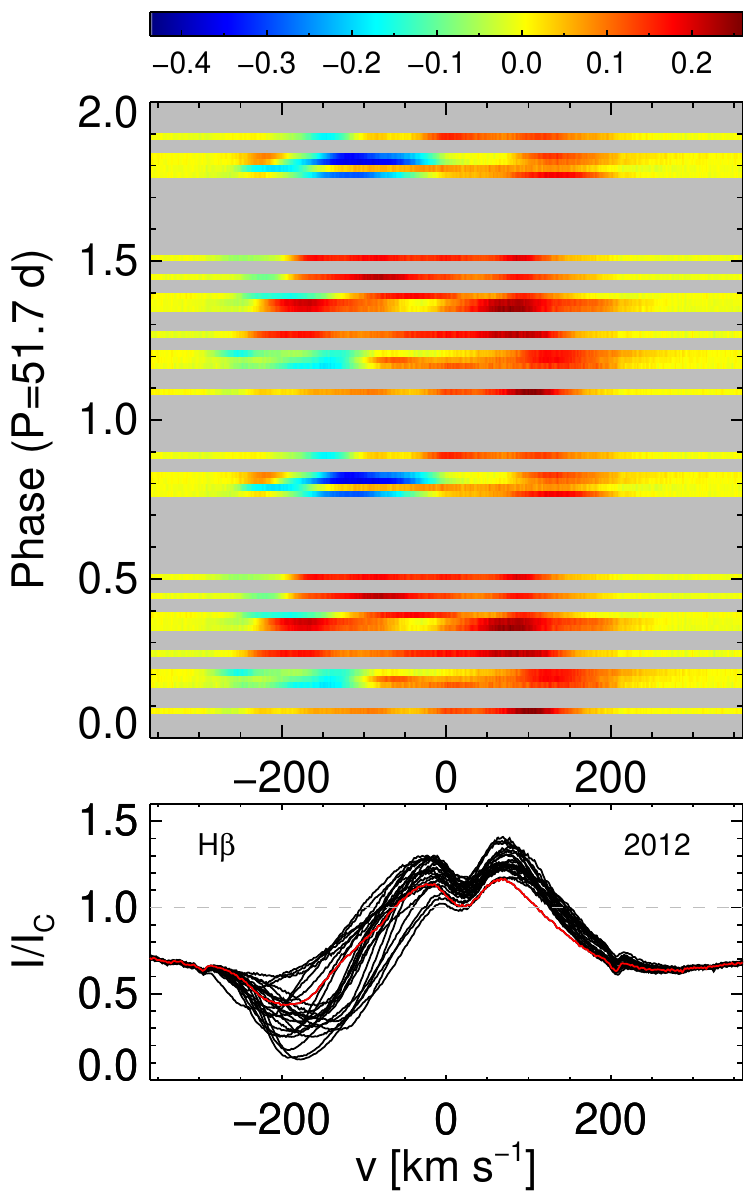}
\includegraphics[width=0.26\textwidth]{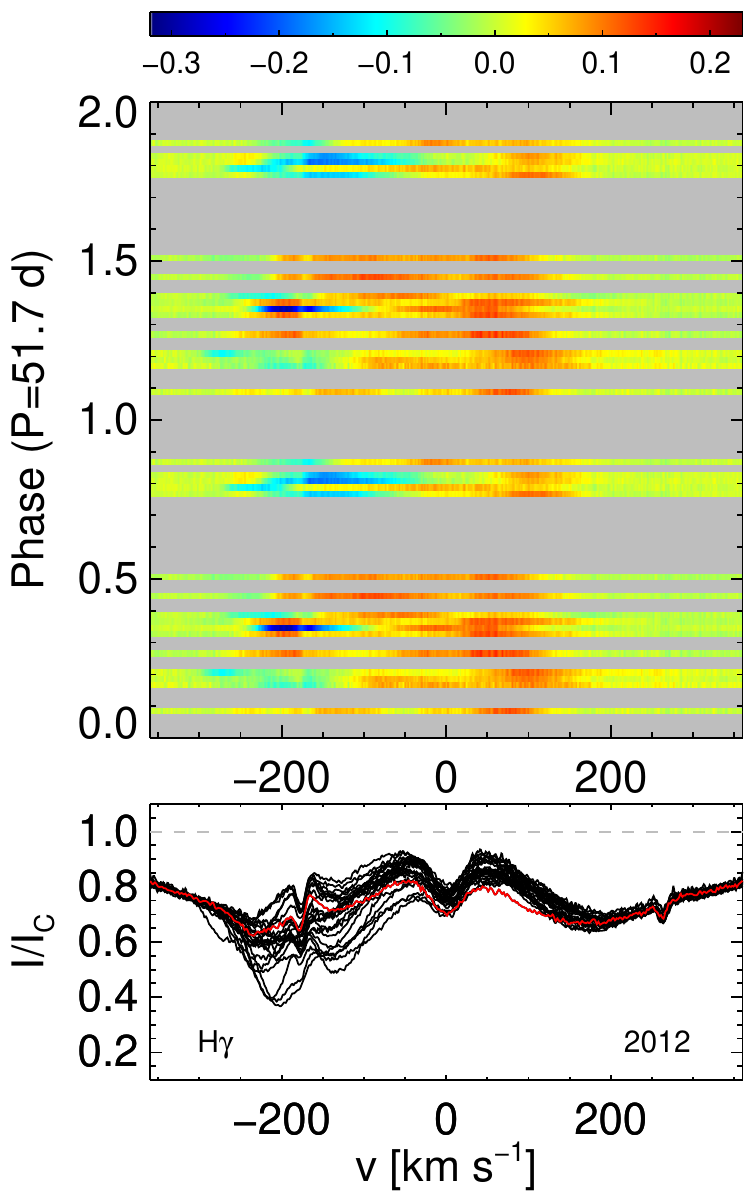}
\caption{
Dynamical spectra for the data acquired in 2012.
Observed Narval 2012 overplotted line profiles (bottom panels) of H$\alpha$
(left), H$\beta$ (middle), and H$\gamma$ (right), and dynamical spectra of
differences between the individual hydrogen line profiles and the profile with
the lowest core emission highlighted by the red colour. The phases are based
on the period determined in Sect.~\ref{sec:Bz}, $P_{\rm rot}=51.7$\,d.
}
\label{fig:Hdyndiff12}
\end{figure*}

\begin{figure*}
\centering
\includegraphics[width=0.26\textwidth]{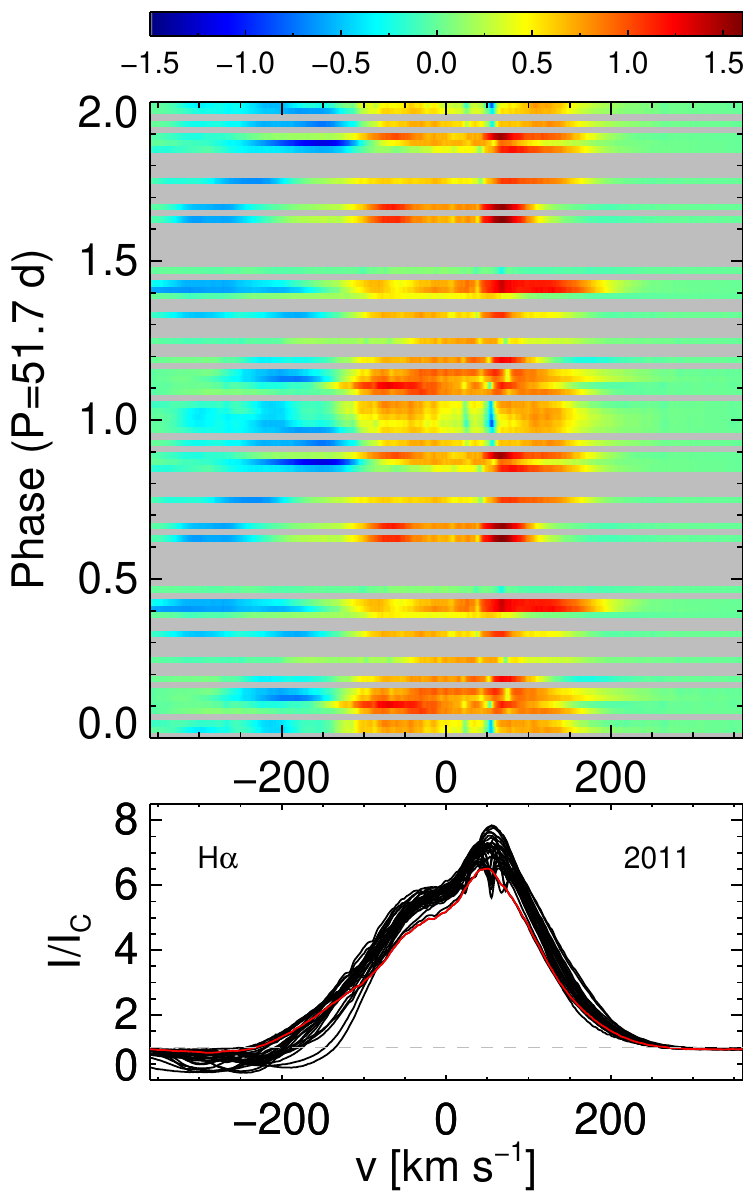}
\includegraphics[width=0.26\textwidth]{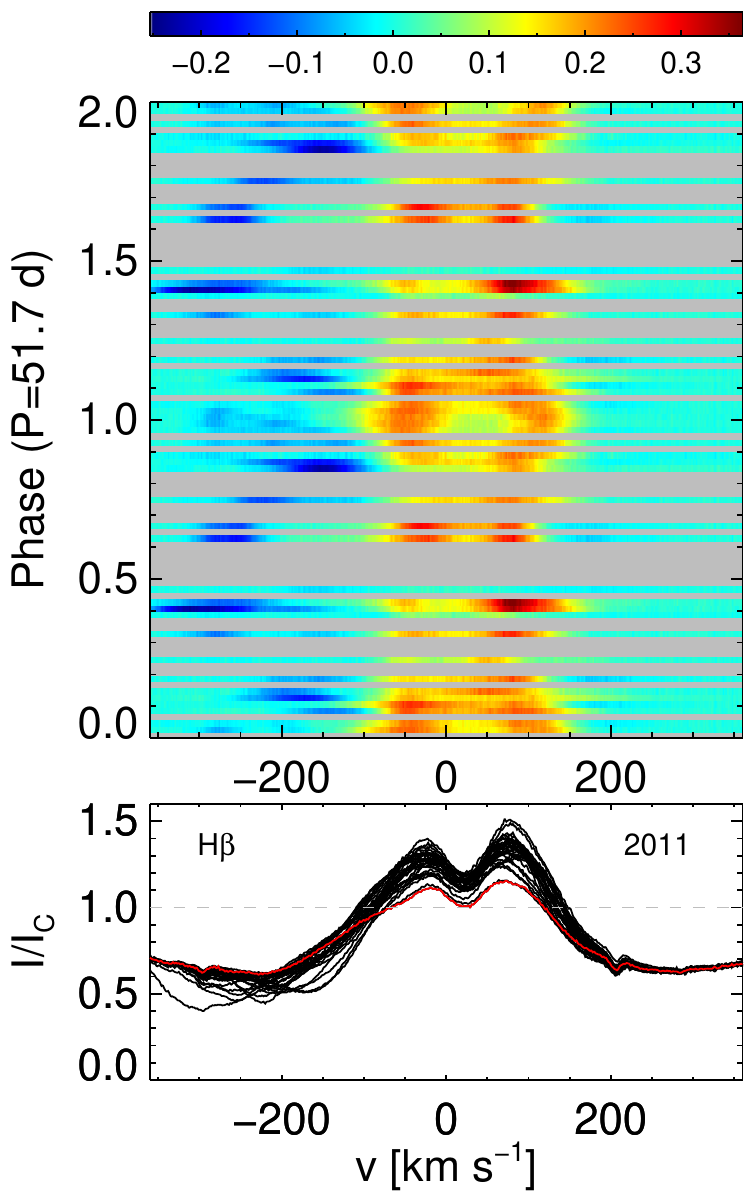}
\includegraphics[width=0.26\textwidth]{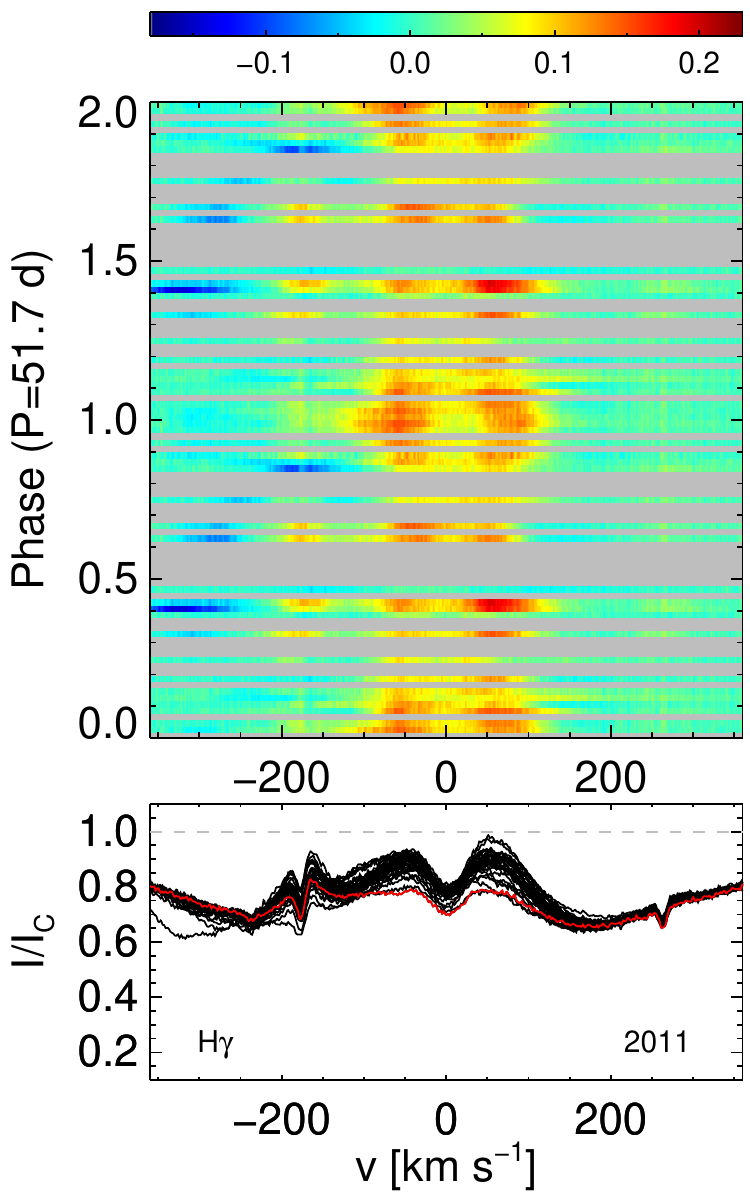}
\caption{
As Fig.~\ref{fig:Hdyndiff12}, but for 2011.
}
\label{fig:Hdyndiff11}
\end{figure*}

Previous studies of highly variable emission line profiles observed in 
magnetic main-sequence stars demonstrate the great usefulness of dynamical 
plots to extract important information on the spectral variability over the 
rotation period
\citep[e.g.][]{Kueker2024}.
Therefore, we used dynamical spectra constructed for hydrogen emission lines 
observed in 2011 and 2012. To achieve the highest contrast in these spectra, 
we used differences between individual hydrogen line profiles and the 
profiles with the lowest core emissions.
Even with our refined period, for the data acquired in 2012 the dynamical 
spectrum presented in Fig.~\ref{fig:Hdyndiff12} does not show much structure 
because of the low number of observations and rather large gaps in the phase 
coverage. Remarkably, as shown in Fig.~\ref{fig:Hdyndiff11} based on the 
data acquired in 2011, we detect a clear ringlike magnetospheric structure 
centred around the phase 0 corresponding to the best visibility of the 
positive magnetic pole. This is the first snapshot of a magnetosphere  around a 
Herbig Ae/Be star. Taking into account the inclination, $i$, and the magnetic 
field geometry with the obliquity, $\beta$, of about $83^{\circ}$ discussed 
in Sect.~\ref{sec:Bz}, we may also expect to see a magnetospheric ring of 
lower intensity around the negative magnetic pole at the phase 0.5, but no 
definite conclusion can be deduced due to the low number of spectra at this 
rotational phase.

The dynamical spectrum based on the 2011 Narval spectra phased with the period 
reported by
\citet{Alecian2013b}
is presented in Fig.~\ref{afig:Hbeta39}. While hinting at a ringlike 
structure in the variability pattern, it does not reveal this feature in a 
prominent way.

\subsection{Metal lines}\label{sec:metal}

Because metal lines in the spectra of HD\,190073 are strongly contaminated 
by emission, for our variability study we selected a few lines of special 
interest. One of them, the emission \ion{O}{i} line at 8446\,\AA{} shows a
significantly fainter profile in 2012 compared to the profile recorded in 
2011. Moreover, in 2006 the emission part of the \ion{O}{i} line profile showed 
three emission peaks of similar intensity 
\citep{Aarnio2017},
but in 2011 and 2012 the peak in the middle is much lower than the other two 
peaks. The dynamical spectra and the overplotted profiles for 2011 and 2012 
are presented in Fig.~\ref{afig:Odyndiff12}. The character 
of the variability is not very clear, but it is obvious that in 2011 the 
emission strength was lower at phase 0 at the best visibility of the positive 
magnetic pole and stronger closer to the negative pole. 

It appears that the variability of the intensity of the metal absorption lines 
is similar to that of the \ion{O}{i}\,8446\,\AA{} line with equivalent 
widths stronger around the negative pole. As an example, we show in 
Fig.~\ref{afig:Feoverplots} the overplotted line profiles, the corresponding 
periodogram, and the phase curve for the \ion{Fe}{ii}\,6149\,\AA{} line 
observed in 2012 and not strongly affected by emission. This line is rather 
weak and the intensity changes are rather small. From the periodicity search 
we obtain $P_{\rm rot}=51.1\pm0.6$\,d, similar to the period determined from 
magnetic field measurements. No significant period has been detected in the 
observations acquired in 2011.

\subsection{The \ion{He}{i}\,10\,830\,\AA{} triplet}

\begin{figure}
\centering
\includegraphics[width=0.16\textwidth]{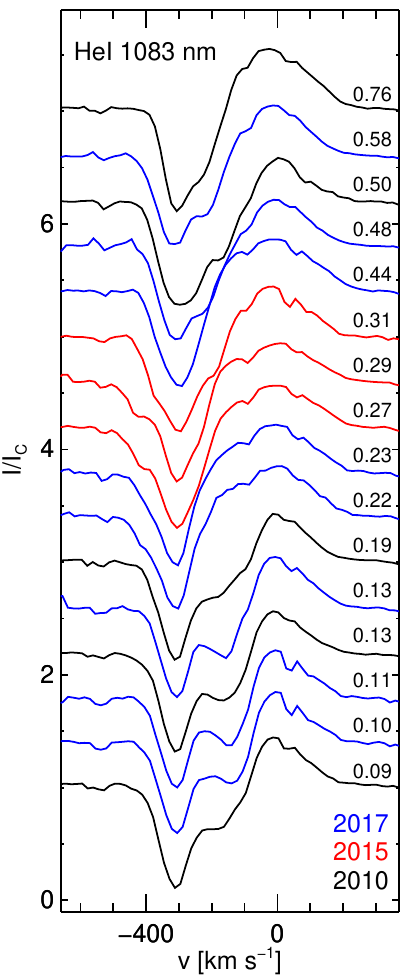}
\includegraphics[width=0.25\textwidth]{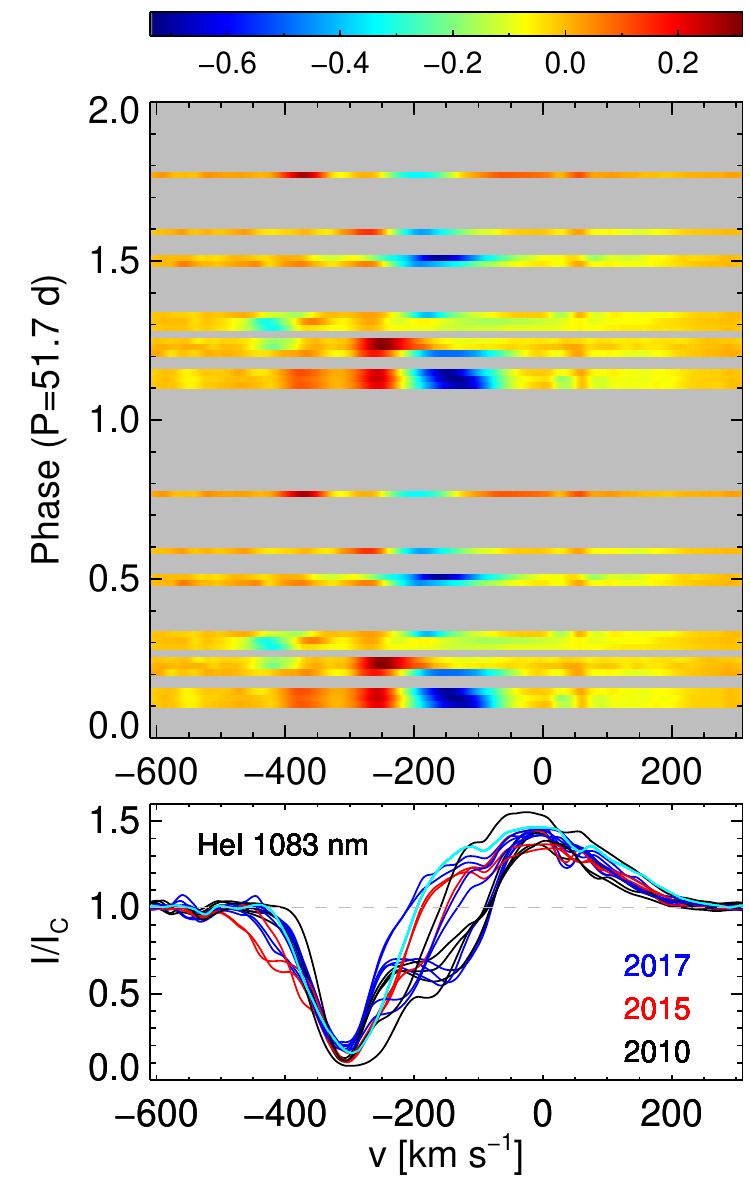}
\caption{
Variability of the \ion{He}{i}\,10\,830\,\AA{} triplet.
{\it Left:} Line profiles of the \ion{He}{i}\,10\,830\,\AA{} triplet observed
in the X-shooter spectra collected during 2010 (black), 2015 (red), and 2017
(blue). The phases indicated on the right are based on the rotation period
of 51.7\,d determined in Sect.~\ref{sec:Bz}.
{\it Right:} Dynamical spectrum for the \ion{He}{i}\,10\,830\,\AA{} triplet.
It is constructed using differences between individual He profiles
and the profile  observed at the rotation phase 0.44 showing the least
complex structure.
}
\label{fig:HeI}
\end{figure}

As mentioned in Sect.~\ref{sec:obs}, the \ion{He}{i}\,10\,830\,\AA{} triplet
is an important diagnostic for tracing the structure and  kinematics of 
accretion and outflow in young stars. Due to the nearly face-on disk 
orientation, we do not expect to detect significant accretion signatures in 
HD\,190073, but should be able to trace the structure of the outflowing wind.
On the left side of Fig.~\ref{fig:HeI}, we show the P\,Cygni profiles of 
this triplet observed in the years 2010, 2015, and 2017, covering the rotation 
phases from 0.1 to 0.76. The blueshifted absorption component extends in some
spectra to more than $-400$\,\kms{}, suggesting the presence of a strong 
stellar wind. It is intriguing that the blueshifted absorption component of 
the line profile shows a prominent feature in the red wing that 
persists over the rotation phase ranges 0.1--0.2 and 0.5--0.8 and decreases 
the contribution of the emission component, probably by strong wind screening.
Between these phase ranges, we see an increase in a weaker feature in the 
blue wing of the absorption component. To better understand the character of 
the variability, we present on the right side of Fig.~\ref{fig:HeI} a
dynamical spectrum of differences between individual He profiles and the 
profile observed at the rotation phase 0.44, showing the least complex 
structure. We can see that the feature in the blue wing appears only
in phases when the red feature disappears. Since we do not have a good 
rotation phase coverage, we can only speculate that the observed variability 
is related to the rotation modulation of the stellar magnetosphere. Without 
future near-infrared  magnetic field measurements using this line, it remains 
unclear whether the emission component of the \ion{He}{i}\,10\,830\,\AA{} 
triplet comes from the magnetosphere or from magnetically controlled 
accretion onto the star, with the accretion streams guided by the magnetic
field lines of the magnetosphere, or from both. Inverse P\,Cygni profiles 
with redshifted absorption reaching several hundred kilometres per second are
usually considered as evidence for magnetospheric accretion. These are not 
observed for HD\,190073 at any rotational phase.

\section{Numerical simulations of the magnetosphere of HD\,190073}\label{sec:MHD}

Knowledge of the fundamental parameters and the dipole field strength of 
HD\,190073 allows us to characterise its magnetosphere. We ran a series 
of 2D MHD simulations of a line-driven wind originating from HD\,190073 
using the {\sc Nirvana} MHD code 
\citep{Nirvana1, Nirvana2}, 
with a setup similar to that of 
\citet{Kuker2017}.  
The initial magnetic field is a dipole rooted in the star. We used spherical 
polar coordinates ($r$,$\theta$) assuming symmetry with respect to the 
magnetic field axis; to be specific, the co-latitude, $\theta$, is measured 
against that axis. The simulation box is a spherical shell, with the inner 
boundary at the stellar surface and the outer boundary at $20\,R_*$. In the 
meridional direction, the box spans the full interval [0,$\pi$], that is, no 
symmetry with respect to the equatorial plane is assumed. 

The main difference between the model of 
\citet{Kuker2017} 
and the one used here is that we solved the equation for the gas energy as 
well as the equations for the magnetic field and the gas motion; in other 
words, the assumption of an isothermal gas has been dropped. We used the 
cooling curve by 
\citet{Schure2009} 
to account for radiative cooling of the gas for temperatures up to $10^8$ K. 
The cooling rate is
\begin{equation}
 L=\int{n_e n_H \Lambda_{N}(T)dV},
\end{equation}
where $n_e$ and $n_H$ are the number densities of electrons and protons, 
respectively. The cooling curve, $\Lambda_N(T)$, for solar metallicity is 
tabulated in 
\citet{Schure2009}.
For temperatures above $10^8$\,K we used the cooling law from 
\citet{Rybicki1979}
\begin{equation}
  \Lambda = L_0 T^{1/2} (1+4.4\times 10^{-10} T ),
\end{equation}
where the constant, $L_0$, has been adjusted to produce a smooth transition 
from the 
\citet{Schure2009} 
cooling curve.

\begin{table}
\begin{center}
\caption{Parameters used in the simulations.}
\begin{small}
\begin{tabular}{cccccc}
\hline\hline\noalign{\smallskip}
$M_*$       & $R_*$       & $T_{\rm eff}$ & $B_0$ & $\dot{M}$              & $\varv_\infty$ \\
$[M_\odot ]$ & $[R_\odot ]$ & $[$K$]$ & $[$G$]$ & $[M_\odot$\,yr$^{-1}]$ & $[$\kms$]$ \\
\hline\noalign{\smallskip}
$6$         & $9.68$      & 9750          & 250   & $10^{-9}$              & 1400 \\
\hline
\end{tabular}\label{tab:star}
\end{small}
\end{center}
\end{table}

The physical parameters used in the simulations are listed in 
Table~\ref{tab:star}. The stellar parameters were adopted from 
\citet{GuzmanDiaz2021}. 
\citet{Catala2007}
found a value of $1.4 \times 10^{-8} M_\odot$\,yr$^{-1}$ for the mass loss 
rate but stated that this value is an upper limit. We therefore adopted 
a lower value of $10^{-9} M_\odot$\,yr$^{-1}$. The value for $\varv_\infty$ 
results from an isothermal simulation run without a magnetic field. We note 
that values for the mass loss rate and terminal velocity refer to the start 
solution, not the evolved state.

We started the simulations with a spherically symmetric isothermal wind with 
constant mass flow as in the isothermal case, with a temperature of $10^5$\,K
rather than the effective temperature of the star. To avoid a discontinuity,
we also used this value on the inner boundary. The gas velocity is given by
\begin{equation}
 \varv_r = \varv_\infty \left(1-\frac{R}{r}\right)^\beta,
\end{equation}
where R is the stellar radius, $\varv_\infty$ the terminal velocity from an 
isothermal model without a magnetic field, and $\beta<1$ a free parameter. In 
the current case, we have used a value of $0.8$.

\begin{figure}
 \begin{center}
 \includegraphics[width=0.35\textwidth]{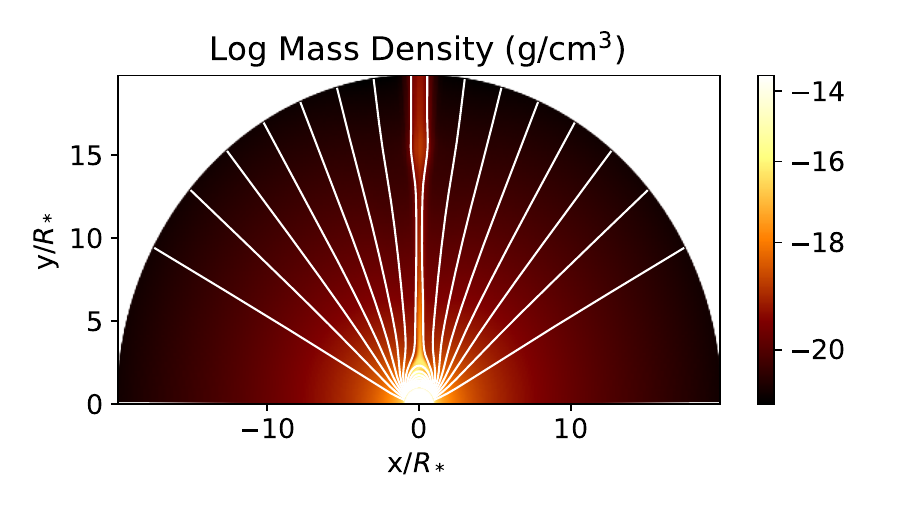}
 \includegraphics[width=0.35\textwidth]{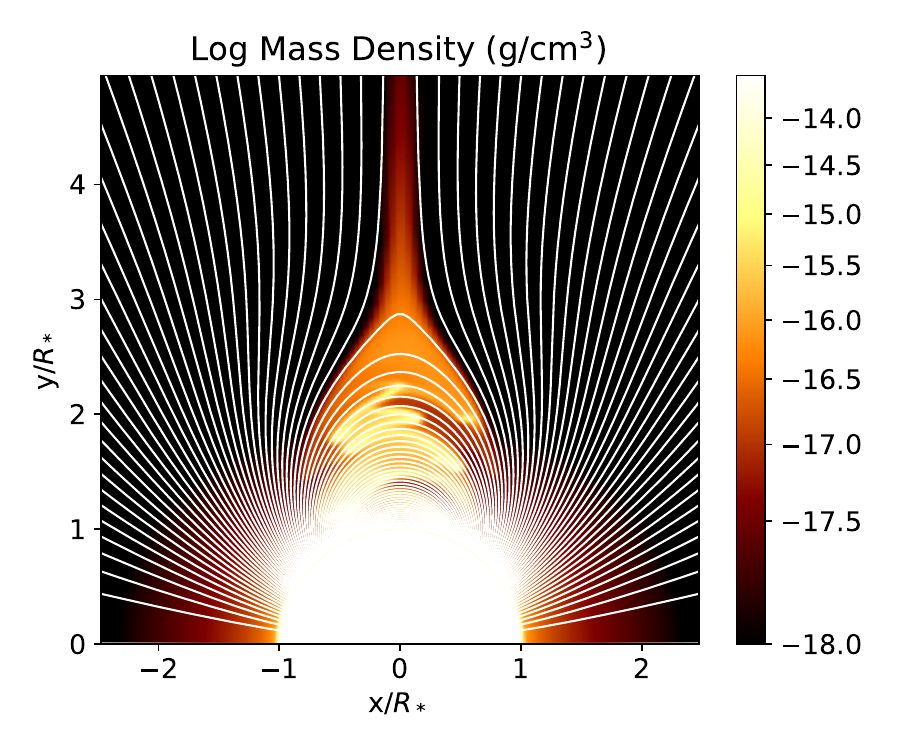}
 \end{center}
 \caption{
Density distribution and magnetic field geometry for our model of
HD\,190073. We used a mass loss rate of $10^{-9} M_\odot$\,yr$^{-1}$ and a 
polar magnetic field strength of 250~G. The white lines represent magnetic 
field lines, the underlying colour contour plot the mass density.
{\it Bottom:} Zoom-in of the magnetosphere.}
 \label{fig:density}
\end{figure}

\begin{figure}
\includegraphics[width=0.24\textwidth]{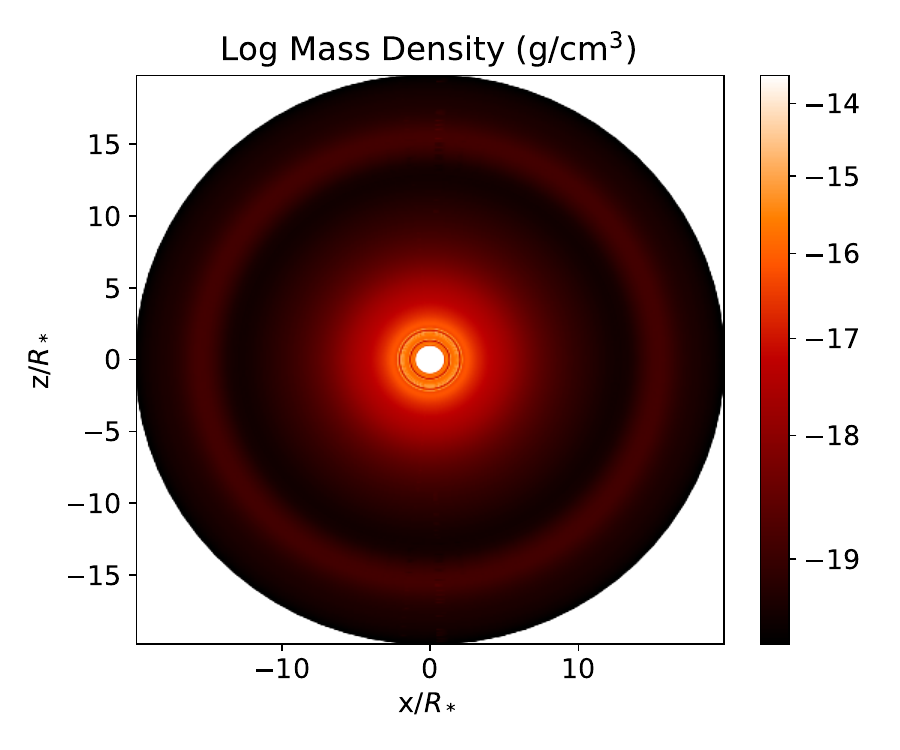}
\includegraphics[width=0.24\textwidth]{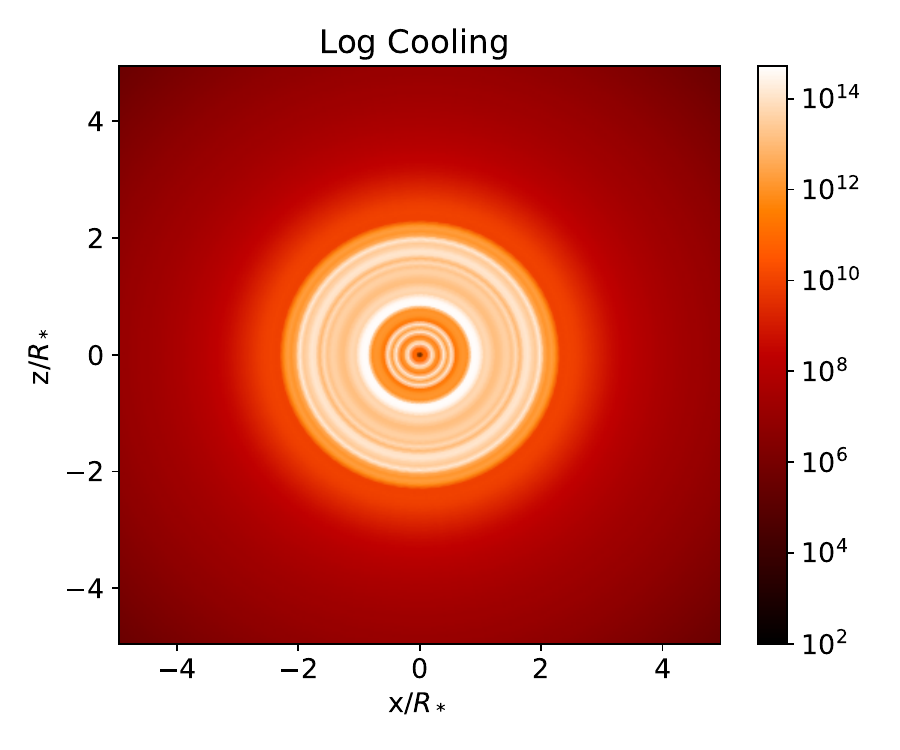}
\caption{
Mass density distribution in the equatorial plane (left)
and vertically integrated cooling function (right).
} 
    \label{fig:rhopole}
\end{figure}

While the gas flow was allowed to evolve freely, the magnetic field was 
kept fixed on the surface of the star. The system evolves rapidly in the 
initial phase of the run but reaches a quasi-steady state after a simulation 
time of about $2\times 10^5$\,s. Figure~\ref{fig:density} shows the mass 
density and the magnetic field from a snapshot of the evolved system, taken 
at $t=5.8 \times 10^5$ s. The dipole axis is in the $x$ direction. The 
magnetic field is still close to the original dipole near the star at low 
(magnetic) latitudes. Field lines originating at high latitudes have opened 
up and a current sheet has formed in the equatorial plane. Outflowing gas 
from the star is trapped in the region of closed field lines and forms a 
magnetosphere. At the outer edge of the magnetosphere, where the field lines 
open up, gas flows originating from mid-latitudes converge and merge in the 
equatorial plane and a disk forms. The top panel of the figure shows an 
increased thickness of the disk near the outer boundary. This feature, which 
is also visible on the left side of Fig.~\ref{fig:rhopole}, is transient, as 
it moves outwards and leaves the simulation box through the outer boundary. 
This ring of outwards-moving gas originates from close to the outer edge of 
the magnetosphere and is accompanied by reconnection in the equatorial 
plane. In the magnetosphere, the gas moves inwards along the field lines in 
a kind of zigzag motion, as can be seen in the bottom panel of 
Fig.~\ref{fig:density}.

\begin{figure}
\includegraphics[width=0.24\textwidth]{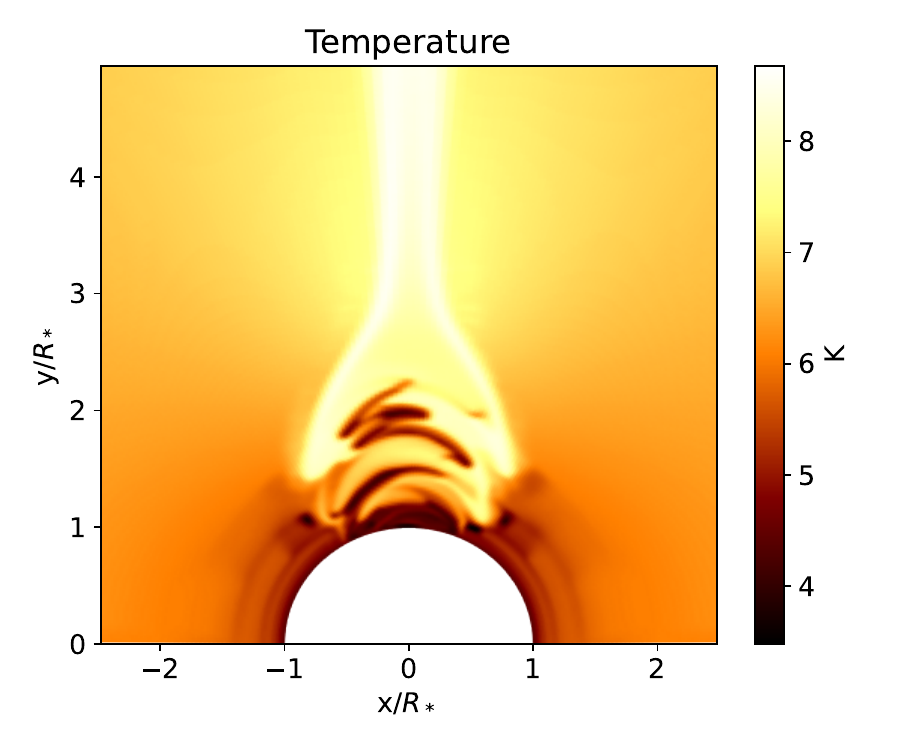}
\includegraphics[width=0.24\textwidth]{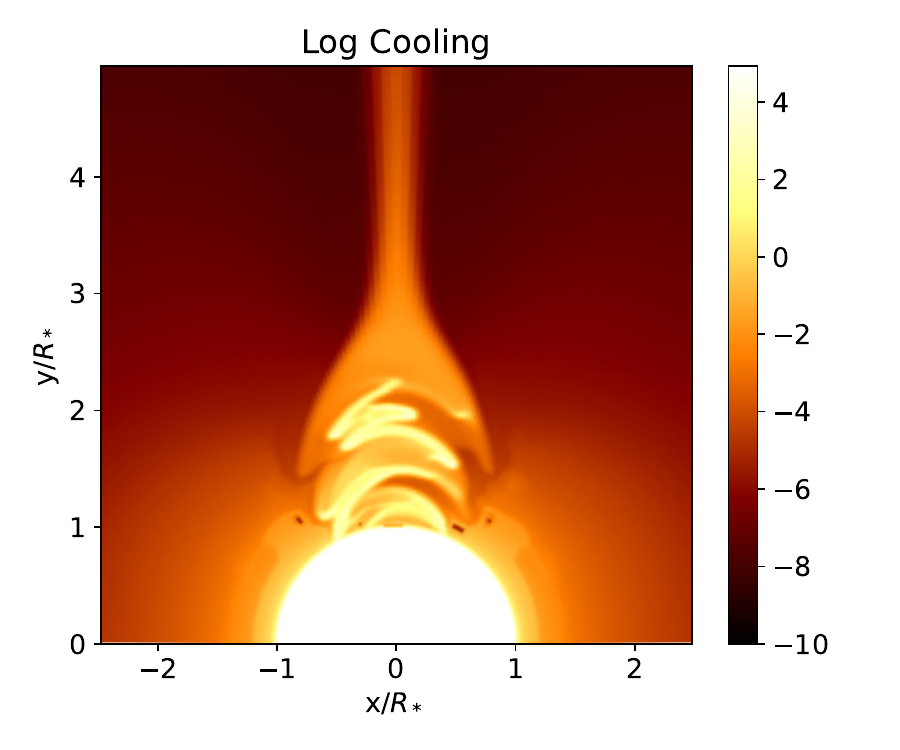}
\caption{
Temperature distribution (left) and cooling function (right) for 
the same snapshot as Fig.~\ref{fig:density}.
}
\label{fig:temperature}
\end{figure}

The gas density in the magnetosphere is much higher than in the wind and 
disk regions of open field lines. The density distribution does not look 
fundamentally different from the isothermal case, but the temperature 
distribution does. The temperature of the outflowing gas is much higher than 
the effective temperature of the star, as shown in the left panel of 
Fig.~\ref{fig:temperature}. It is particularly high in the disk and in the 
low-density parts of the magnetosphere. The high-density parts of the latter 
are much cooler though, closer to the effective temperature of the star.

The adopted value for the dipolar magnetic field strength was slightly higher 
than the value derived from the observations but lies within the error 
margin. A lower value of 222\,G will lead to a slight reduction in the 
confinement parameter if the values of the mass loss rate and terminal 
velocity remain the same. Given the uncertainty in the latter two 
quantities, a change of the dipole magnetic field strength to 222\,G has 
only a minor impact on the results. 

We find a smaller magnetosphere than predicted from isothermal models. 
The extent of the magnetosphere is given by the radius 
\citep{Uddoula2008}
\begin{equation}
R_c\approx R_*+0.7 (R_A-R_*),
\end{equation} 
where the Alfv\'en radius $R_A$ is given by
\begin{equation}
\frac{R_A}{R_*} \approx 0.3 +\eta_*^{1/4}
\end{equation}
and the confinement parameter, $\eta_*$, by
\begin{equation}
 \eta_*=\frac{B^2_{\rm eq} R^2_*}{\dot{M} \varv_\infty}.
\end{equation}

With the values listed in Table \ref{tab:star}, we find 
$\eta_* = 800$, $R_A / R_* = 5.6$, and $R_C / R_* =4.2$ for the isothermal 
case. Using $B_0=222\,G$, we would get $\eta_* = 630$, $R_A / R_* = 5.3$, 
and $R_c / R_* = 4.0$.

While causing the observed variability, the rotation is too slow to be 
dynamically important. The impact of rotation is measured by the parameter
\begin{equation}
 W = \frac{V_{\rm rot}}{V_{\rm orb}},
\end{equation} 
where $V_{\rm rot}$ is the equatorial rotation velocity of the star at its 
(rotational) equator and $V_{\rm orb}$ the Keplerian velocity at the same 
point. We find $W=0.03$, which means that the magnetosphere is a dynamical 
magnetosphere rather than centrifugal. We therefore limit the simulations to 
the nonrotating case.

The effect of including thermodynamics is demonstrated in 
Fig.~\ref{fig:density}, showing closed field lines up to a radius of 3.0 for 
the last closed field line. The disk also appears less pronounced and less 
fractured. Reconnection events leading to the ejection of gas from the 
outer edge of the magnetosphere happen episodically, but less often than in 
the isothermal case, with rings of gas moving outwards. The mass loss rate 
is reduced by an order of magnitude from the start solution, while the 
terminal velocity is increased by a factor of four. The right panel of 
Fig.~\ref{fig:temperature} shows the spatial distribution of the cooling 
function. It reflects the density distribution, with values largest in the 
cool, high-density regions of the magnetosphere. The disk of hot 
outflowing gas is also quite pronounced.

Our simulations are in the $r-\theta$ plane. Figures~\ref{fig:density} and 
\ref{fig:temperature} therefore show an edge-on view of the system. As the 
inclination angle, $i$, is $19.7 \pm 15.8$ degrees and the obliquity angle, 
$\beta$, has a value of $82.9 \pm 6.2$ degrees, this view is a good 
representation of what we observe. For an impression of a pole-on view, we 
have, however, created a three-dimensional data set using axisymmetry. In the 
left panel of Fig.~\ref{fig:rhopole} we show the gas density in the equatorial 
plane. The magnetosphere appears as a ring around the star, while the disk 
forms a fainter halo. The ring shows some substructure, with gaps at certain 
radii caused by the zigzag flow along the field lines in the magnetosphere. 
There is also a ring of enhanced gas density in the outer part of the box. 
This ring is not a static feature but has broken off the inner disk and is 
moving outwards away from the star.

In the right panel of Fig.~\ref{fig:rhopole} we show a zoom-in of the 
vertically (parallel to the dipole axis) integrated cooling function in 
the $r-\phi$ plane. The magnetosphere appears as a bright disk and shows some 
substructure in form of dark and bright rings and a rather sharp outer edge.

Overall, our simulations show a field geometry, gas distribution, and flow 
pattern similar to those from previous simulations that assumed the gas to 
be isothermal at the stellar effective temperature. The inclusion of heat 
transport and cooling, however, leads to a weaker impact of the magnetic 
field. The magnetosphere is smaller than in the isothermal case and the 
region outside the last closed field line shows much less structure. We 
attribute this to the much higher temperatures, which lead to higher gas 
pressure and thus larger values of plasma beta, the ratio 
between gas pressure and magnetic pressure. 

\section{Discussion}\label{sec:dis}

Combining our own and archival spectropolarimetric observations obtained 
between 2012 and 2019 with three different spectropolarimeters, we determine 
for HD\,190073 a rather long magnetic period $P=51.70\pm0.06$\,d. No 
significant period was detected in the data acquired in 2011 between May and 
November. Monitoring of the variability of the H$\beta$ profiles recorded 
from 2005 up to 2023 indicates the presence of a significant distortion of 
the profile shapes, causing striking variability in the depth and position 
of the blueshifted absorption line component and the intensity of the 
emission line component. Apart from the variability of the hydrogen line 
profiles over the rotation cycle, the more substantial changes appear on 
monthly or annual timescales. Since several studies suggest the presence 
of a binary or protoplanetary companion in the system HD\,190073
\citep[e.g.][]{Baines2006, Rich2022,Anilkumar,Ibrahim},
it is possible that the observed profile distortions can be explained by a 
magnetospheric interaction between the magnetic host star and the secondary 
component. 

From our magnetic field measurements in combination with the fundamental 
parameters presented in the literature, we estimate an obliquity angle 
$\beta=82.9\pm6.4^{\circ}$ and a dipole strength $B_{\rm d}=222\pm66$\,G. 
It is remarkable that the two stars with known magnetic phase curves,
HD\,101412 and HD\,190073, have rather long rotation periods, which are
not typical for other Herbig Ae/Be stars, and that in both stars the 
magnetic axis is located close to the stellar equatorial plane. Based on the 
work of 
\citet{Romanova2003}, 
\citet{Hubrig2014} 
suggested that for HD\,101412 with a similar obliquity angle 
$\beta=84\pm13^{\circ}$ and a long rotation period of about 42\,d, many polar 
field lines probably thread the inner region of the disk, while the closed 
lines cross the path of the disk matter, causing strong magnetic braking.

The strong variability of the hydrogen emission lines H$\alpha$, H$\beta$, and 
H$\gamma$ has been studied using dynamical spectra. Due to the rather small 
number of acquired Narval spectra in 2012 July-October, our dynamical 
spectra do not reveal any clear periodic features. However, in the data 
acquired in 2011 May--November, we detect for all three hydrogen emission line 
profiles clear ringlike magnetospheric structures appearing at the rotation 
phase 0 during the best visibility of the magnetic pole. These spectra 
present the first snapshot of a magnetosphere around a Herbig Ae/Be star. Our 
study of the variability of metal lines indicates that their strength varies 
in an opposite way to the strength of the hydrogen emission lines, with 
stronger lines observed closer to the negative magnetic pole. Intriguingly, 
variable absorption features changing their shape over the rotational period 
have been detected in the wings of the blueshifted absorption component of the 
\ion{He}{i}\,10\,830\,\AA{} triplet. We speculate that these features are
related to the rotational modulation of the observed magnetosphere.

For the first time, 2D MHD simulations have been carried out for a Herbig 
Ae/Be star. Using the {\sc Nirvana} MHD code and involving nonisothermal 
gas, our simulations show that the magnetosphere should be rather compact 
with a radius of about three stellar radii, but the wind flow extends over 
tens of stellar radii. With the radius of a disk of 1.14\,au around HD\,190073 
\citep{Ibrahim}, 
the distance between the star 
and the disk is about 25 stellar radii. It is not clear whether an 
interaction between the stellar magnetosphere and the surrounding accretion 
disk would explain the sub-au structure inconsistent with Keplerian motion, 
as 
\citet{Ibrahim} 
report, and their speculation about the presence of an object embedded in 
the inner disk.

As mentioned in Sect.~\ref{sec:introduction}, several studies of HD\,190073 
indicate the possible presence of a lower-mass binary companion very close 
to the host star and a more distant giant planet component, making this 
system a valuable laboratory to study magnetic interaction between a host 
star and its companions. There are currently two ways to confirm the 
reported discoveries of companions. Studying the presence of a companion 
within inner disks of Herbig Ae/Be stars, a resolution below one milliarcsecond 
is required, which may be achievable with a new generation of optical 
interferometric instruments with longer baselines. Current near-infrared 
interferometric surveys  of the inner regions in Herbig Ae/Be stars offer a
spatial resolution of the order of one milliarcsecond. Follow-up polarized 
differential imaging observations with the Gemini Planet Imager are also 
needed to confirm that the detected giant planet and the host star are 
co-moving.

\begin{acknowledgements}

We thank the referee for their comments.
Based on observations made with ESO Telescopes at the La Silla Paranal
Observatory under programme IDs
187.D-0917, 089.D-0383, 097.C-0277, 099.C-0081, 0103.C-0240,
385.C-0131, 095.C-072, and 099.C-0342.
Based on observations collected at the Bernard Lyot Telescope at the Pic du 
Midi de Bigorre, France, which is managed by the Observatoire Midi 
Pyr\'en\'ees, and on observations collected at the Canada-France-Hawaii 
Telescope (CFHT), which is operated by the National Research Council of 
Canada, the Institut National des Sciences de l'Univers of the Centre 
National de la Recherche Scientifique of France, and the University of Hawaii.

\end{acknowledgements}

%
   \bibliographystyle{aa} 
   \bibliography{hd190073} 
%





\begin{appendix} 

\section{Period analysis}

\begin{figure}
\centering
\includegraphics[width=0.23\textwidth]{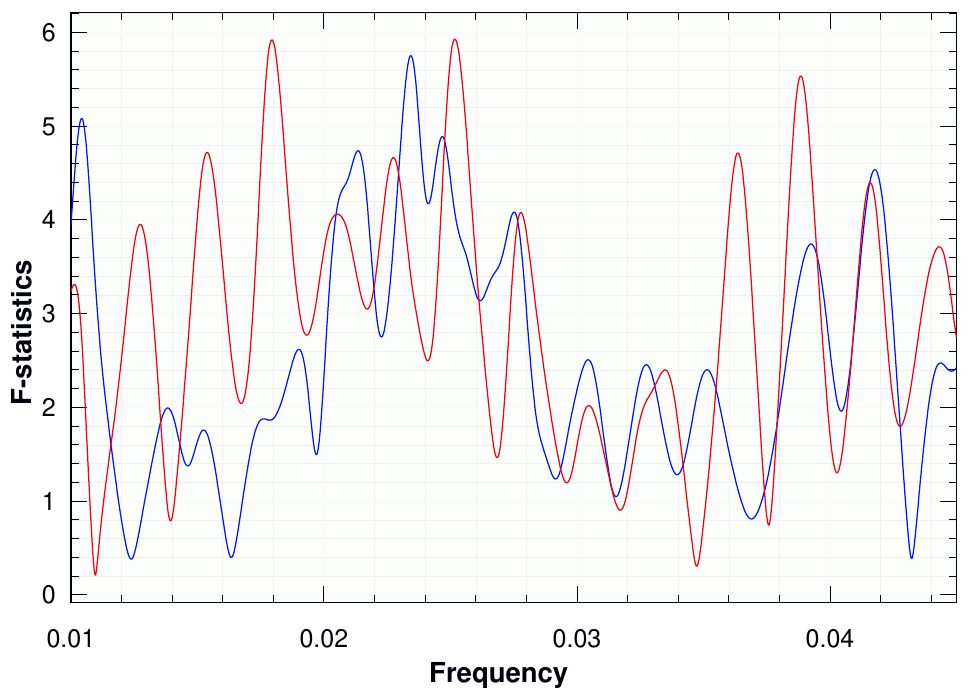}
\includegraphics[width=0.23\textwidth]{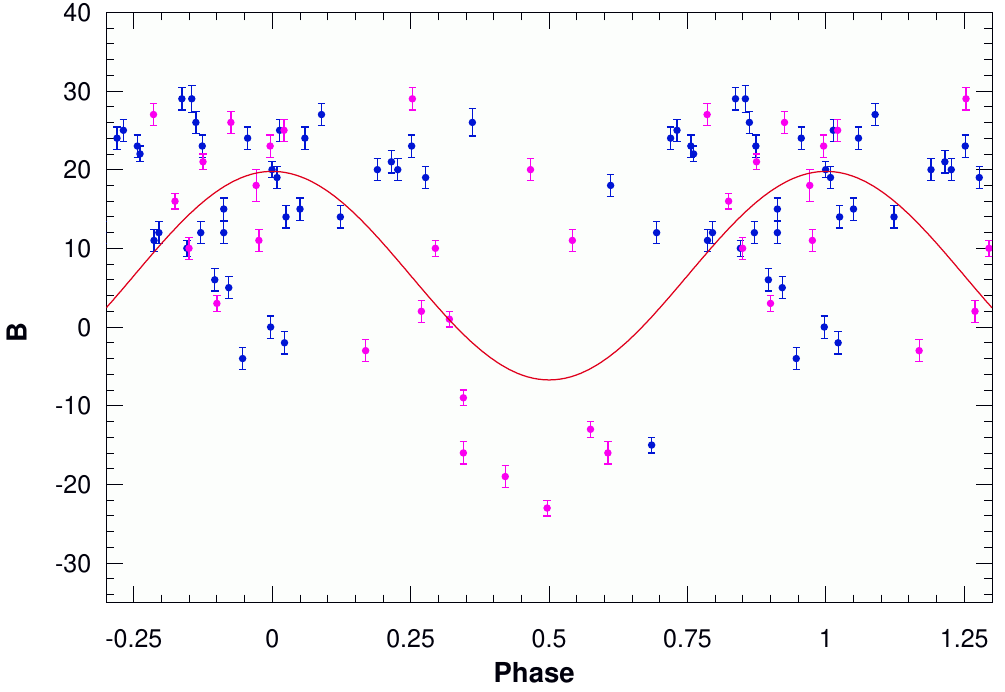}
\caption{
Periodogram based on all Narval measurements from the years 2011
and 2012 (left). The red line is used for the observations and the blue 
line for the window function. $\bz{}$ measurements phased with 
$P=51.6\pm1.1$\,d (right). The sinusoidal fit is shown with the red line. 
}
\label{afig:Bz20112012}
\end{figure}

\begin{figure}
\centering
\includegraphics[width=0.23\textwidth]{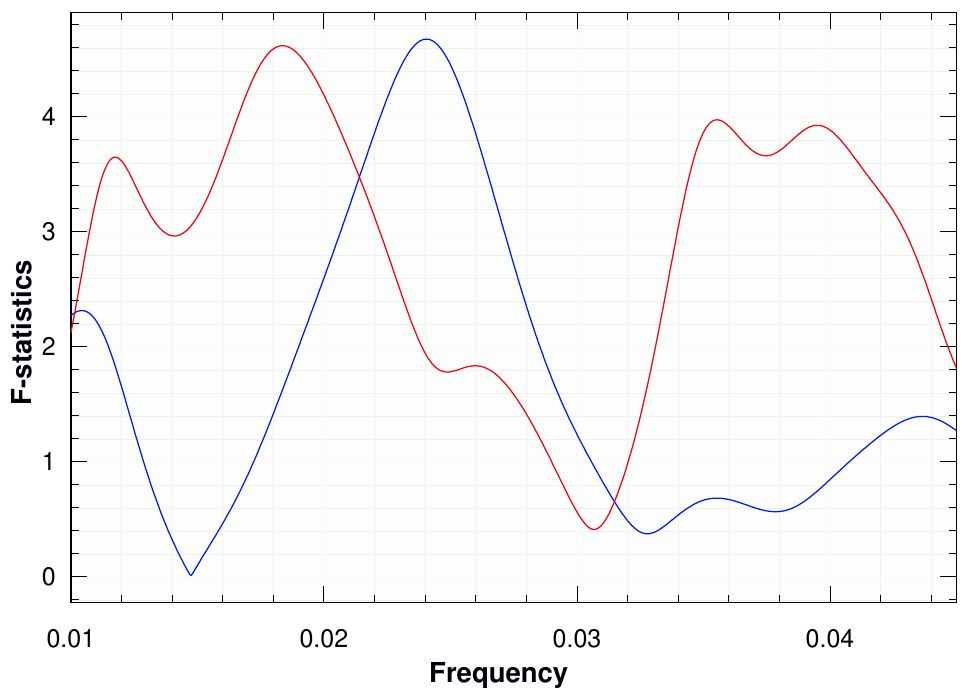}
\includegraphics[width=0.23\textwidth]{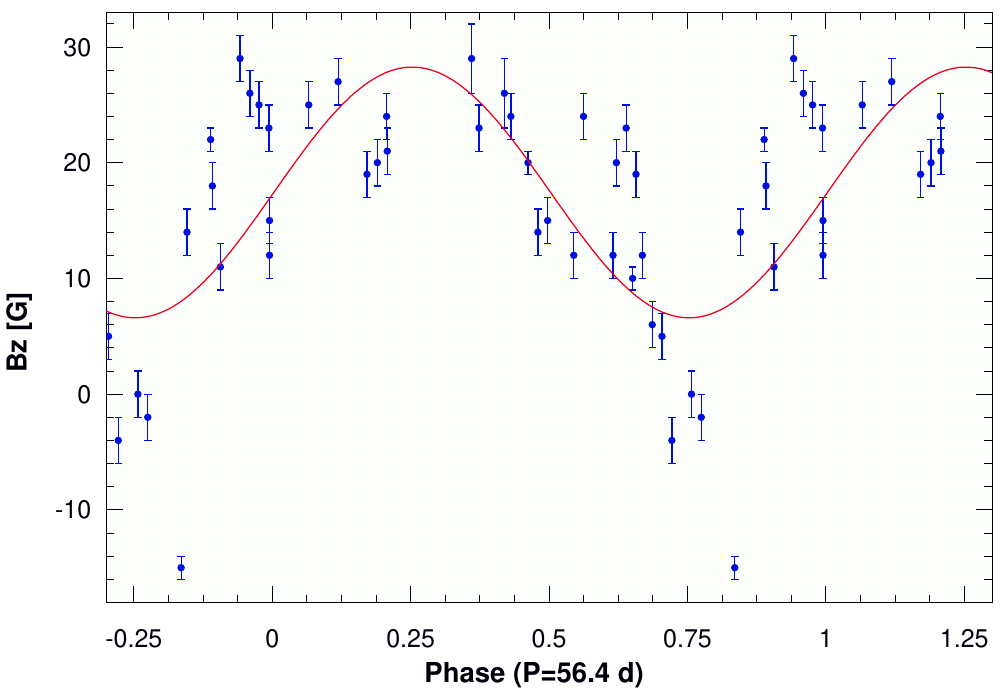}
\caption{
As Fig.~\ref{afig:Bz20112012}, but for the Narval observations in 2011.
The $\bz{}$ measurements are phased with $P=56.4\pm2.7$\,d and  the
sinusoidal fit is shown with the red line.
}
\label{afig:Bz2011}
\end{figure}

\begin{figure}
\centering
\includegraphics[width=0.23\textwidth]{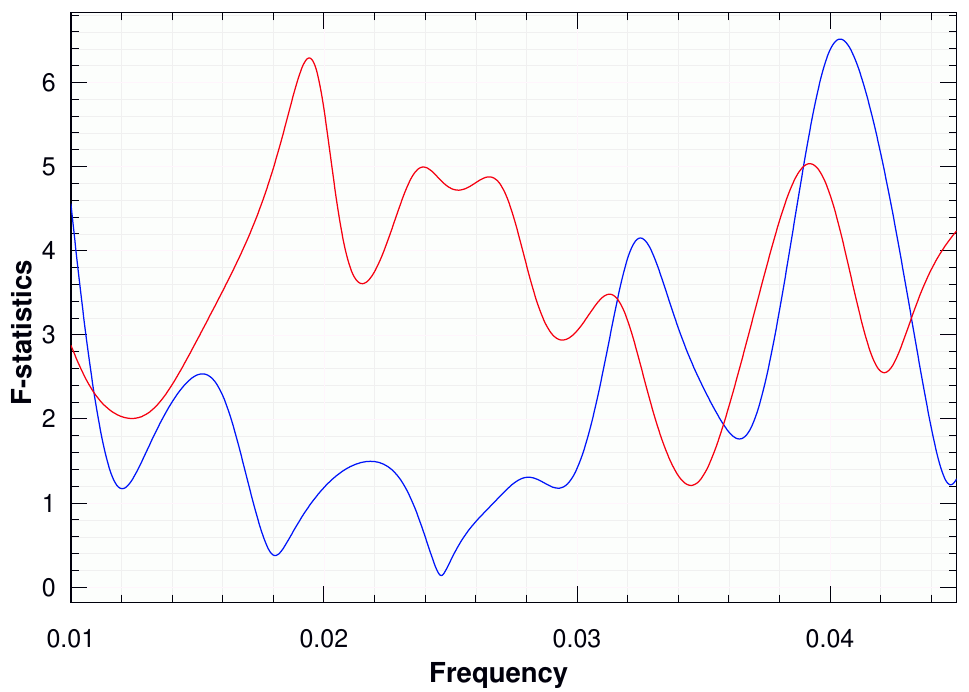}
\includegraphics[width=0.23\textwidth]{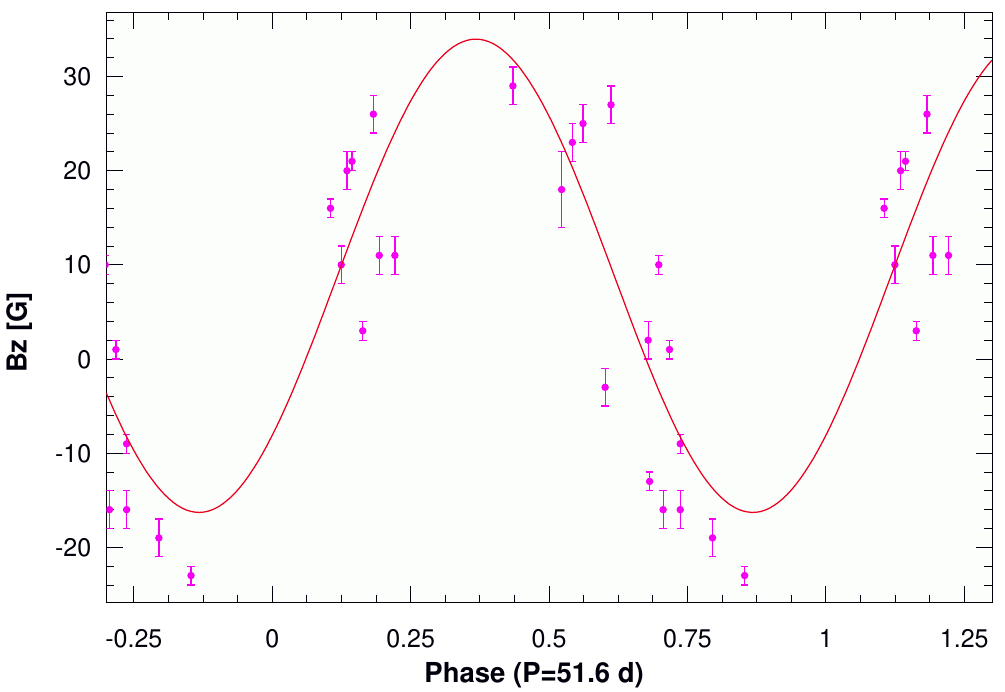}
\caption{
As Fig.~\ref{afig:Bz20112012}, but for the Narval observations in 2012.
The $\bz{}$ measurements are phased with $P=51.6\pm1.1$\,d and the
sinusoidal fit is shown with the red line.
}
\label{afig:Bz2012}
\end{figure}

Our period search based on 2011 and 2012 Narval spectra combined is 
presented in Fig.~\ref{afig:Bz20112012} and separately in 
Figs.~\ref{afig:Bz2011} and \ref{afig:Bz2012}, respectively.

\section{Variability of hydrogen lines}\label{bsec:vari}

\begin{figure}
\centering
\includegraphics[width=0.48\textwidth]{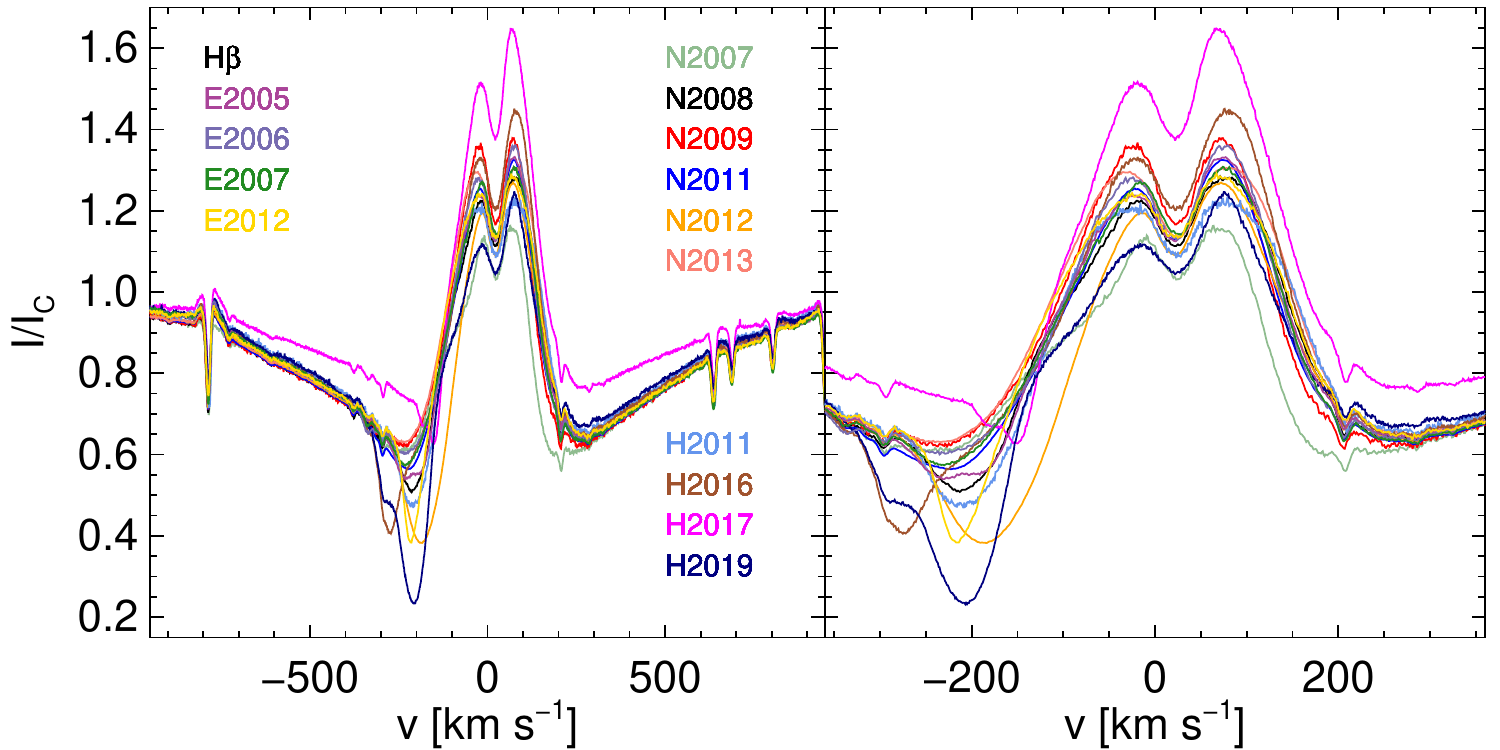}
\includegraphics[width=0.48\textwidth]{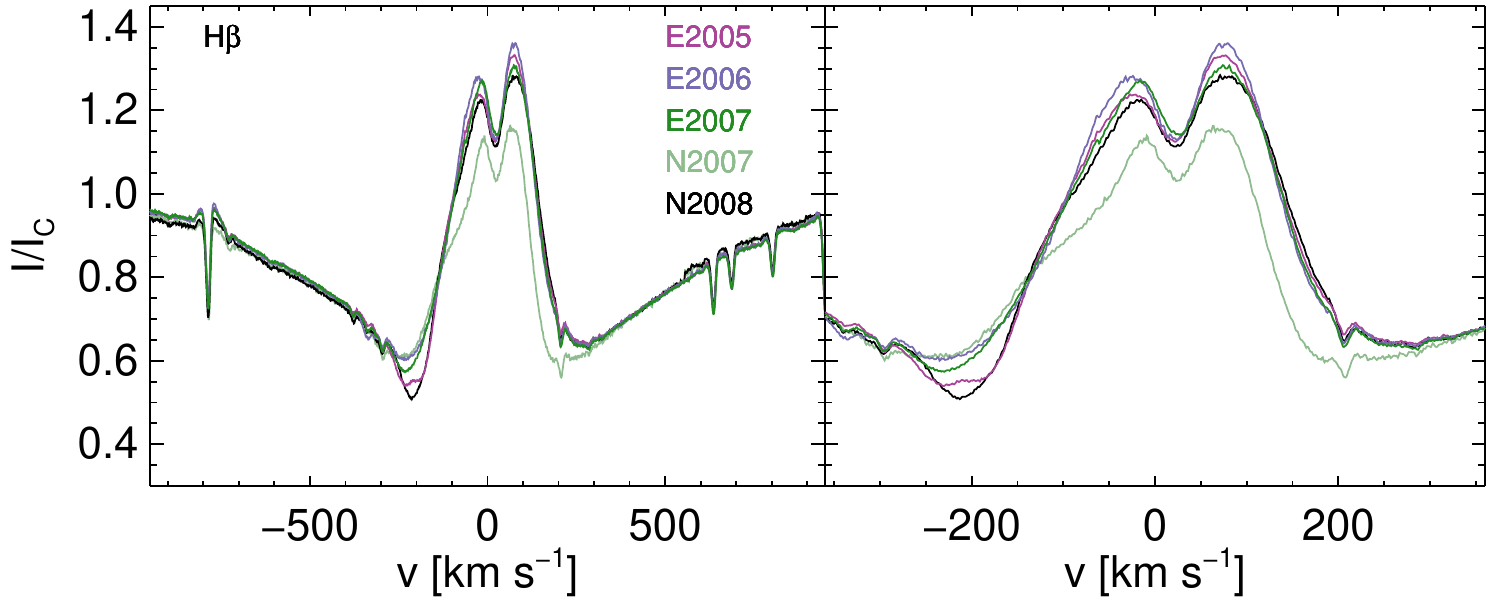}
\includegraphics[width=0.48\textwidth]{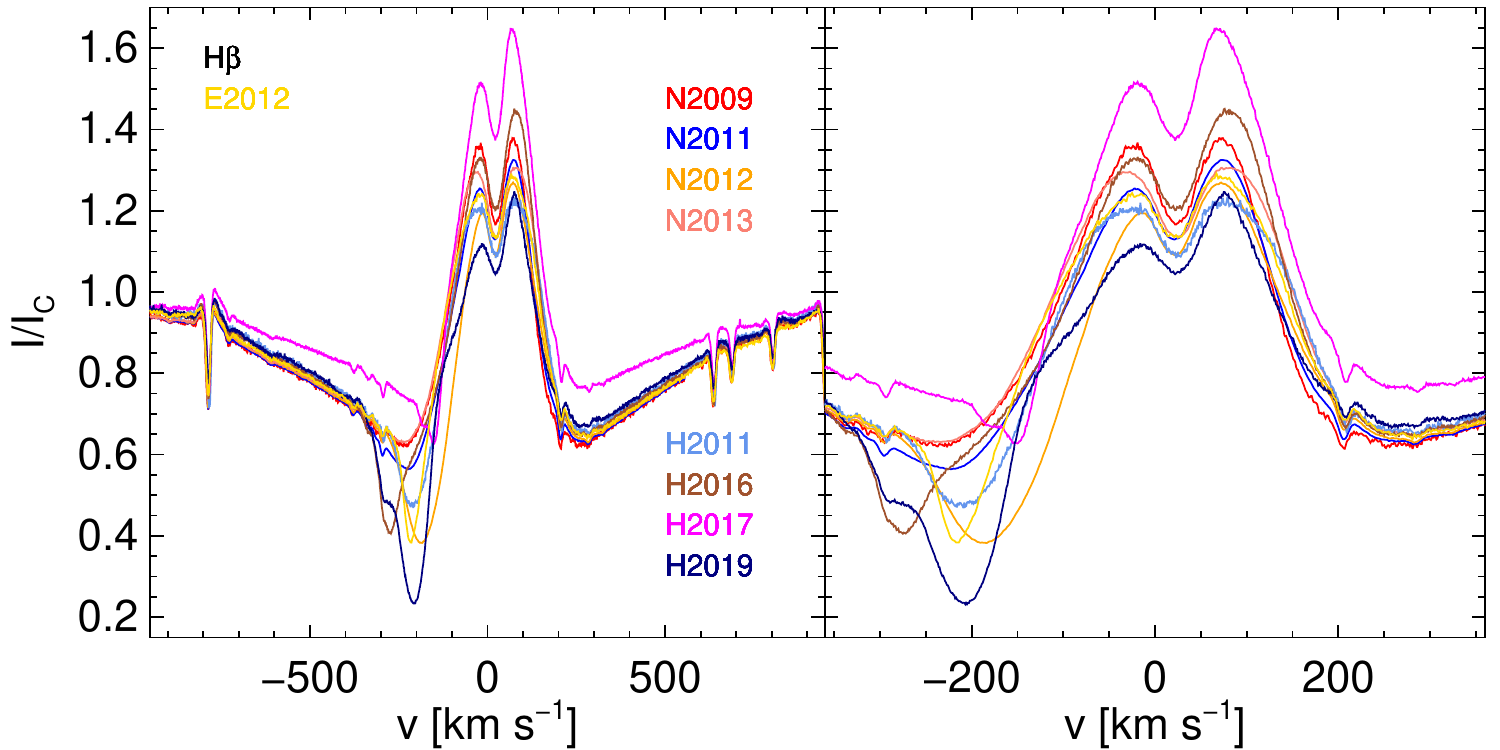}
\caption{
Variability of H$\beta$ profiles over different years.
{\it Top:}The observations have been acquired from 2005 to 2019 using different
instruments: E indicates ESPaDOnS, N Narval, and H HARPS\-pol.
{\it Left:} Larger velocity range to show the full extent of the line profiles.
{\it Right:} Zoom-in to the line core. The profile in magenta with
the strongest emission contribution refers to the  HARPS\-pol observation in
2017 whereas the profile in navy blue colour highlights the HARPS\-pol
observation in 2019. The middle plot shows the profiles until 2008 and the
bottom plot the profiles of the later years.
}
\label{afig:Hbetayr}
\end{figure}

\begin{table}
\centering
\caption{
Spectra used to investigate H$\beta$ profile variability over 
the years. 
}
\label{tab:vari}
\begin{small}
\begin{tabular}{llrr}
\hline\hline\noalign{\smallskip}
Year  & Instrument & Resolution & N.\ of spectra \\
\hline\noalign{\smallskip}
2005  & ESPaDOnS   & 65\,000 &  6 \\
2006  & ESPaDOnS   & &  3 \\
2007  & ESPaDOnS   & &  3 \\
2012  & ESPaDOnS   & &  1 \\
2007  & Narval     & 65\,000 &  2 \\
2008  & Narval     & &  3 \\
2009  & Narval     & &  1 \\
2011  & Narval     & &  37 \\
2012  & Narval     & &  23 \\
2013  & Narval     & &  10 \\
2011  & HARPS\-pol & 115\,000 &  1 \\
2016  & HARPS\-pol & &  2 \\
2017  & HARPS\-pol & &  3 \\
2019  & HARPS\-pol & &  1 \\
\hline\noalign{\smallskip}
2015 & CES         & 14\,000 &  22 \\
2016 & ShAFES      & 28\,000 &  6 \\
2017 & ShAFES      & &  7 \\
2018 & ShAFES      & &  5 \\
2019 & ShAFES      & &  5 \\
2020 & ShAFES      & &  3 \\
2021 & ShAFES      & &  1 \\
2022 & ShAFES      & &  1 \\
2023 & ShAFES      & &  4 \\
\hline
\end{tabular}
\end{small}
\tablefoot{ The first column gives the year, the second the instrument used,
and the third the number of spectra. The high-resolution spectra are listed
in the upper part of the table and the medium-resolution spectra in the
lower part of the table.}
\end{table}

In Fig.~\ref{afig:Hbetayr}, we present the individual and mean (for details,
see Table~\ref{tab:vari}) line profiles of H$\beta$ obtained over different 
years from 2005 to 2019. In the following, we discuss the changes of these 
profiles over time.

The H$\beta$ profiles between 2005 and 2008 show very little variability. 
The line emission component is the weakest in the 2007 Narval profile 
(light green). This profile shows also the strongest red wing absorption 
component among all profiles. The strongest absorption in the blue wing is 
detected in the 2005 ESPaDOnS profile (purple) and the 2008 Narval profile 
(black). After 2008, the profiles showed much more variability. Among all 
the profiles, the emission component is strongest in the 2017 HARPS\-pol 
spectrum (magenta), followed by the 2016 HARPS\-pol profile (brown), and 
the 2009 Narval profile (red). The blueshifted absorption component is 
strongest in the 2019 HARPS\-pol spectrum (navy blue). It also shows one of 
the weakest emission components, comparable to that of the Narval 2007 
profile, albeit the red emission peak is stronger in 2019 than in 2007. 
Both the 2012 ESPaDOnS and Narval spectra (light and dark yellow, 
respectively) show similarly deep absorption line components, but clearly 
weaker than seen in 2019. Also the 2016 HARPS\-pol profile has almost 
equally strong absorption in the blue wing as was detected in 2012. The red 
absorption wing shows very little variability --- the strongest absorption 
is seen in the 2007 Narval profile and the weakest in the 2019 HARPS\-pol 
profile. 

\begin{figure}
\centering
\includegraphics[width=0.48\textwidth]{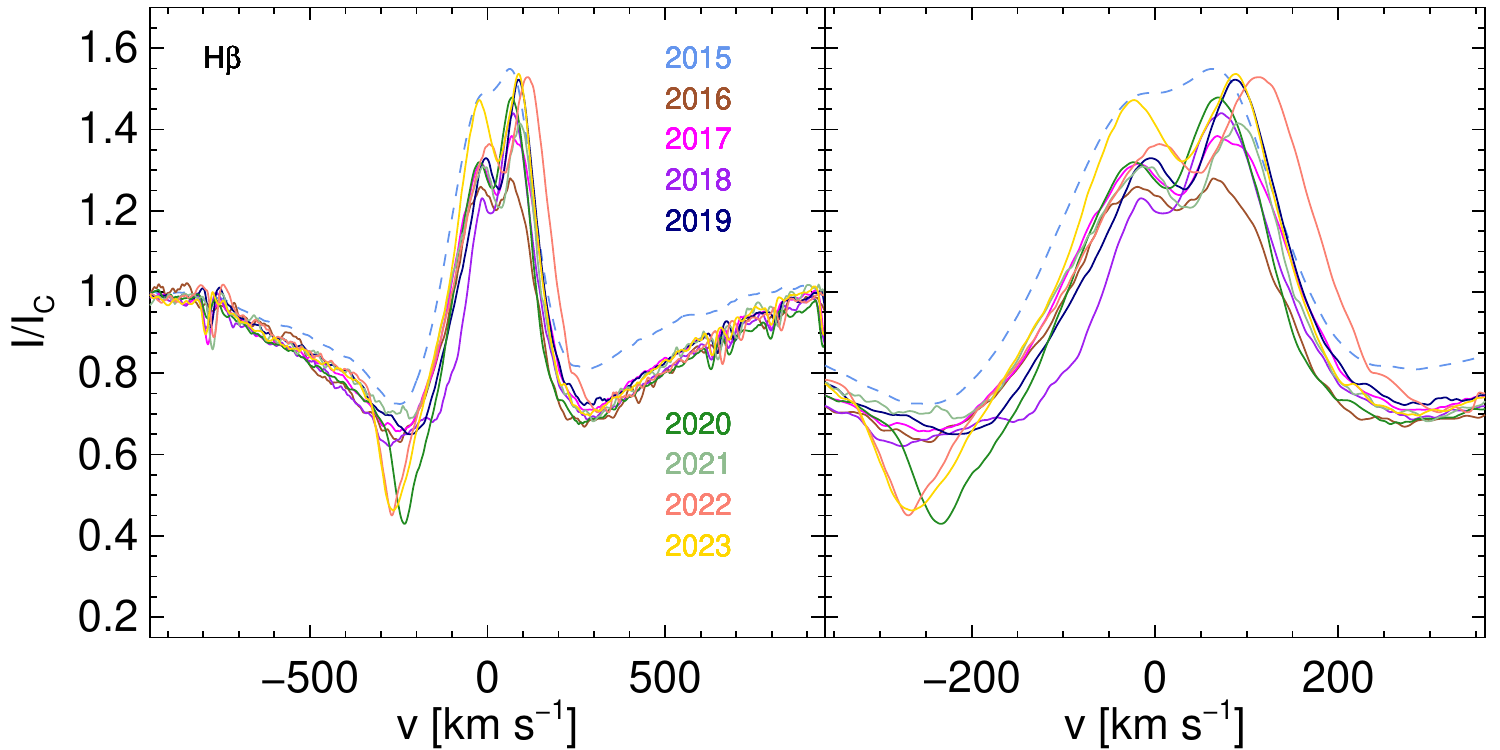}
\caption{
As Fig.~\ref{afig:Hbetayr}, but based on medium-resolution spectra. The plot 
shows the variability of the H$\beta$ profiles over different years from 
2015 to 2023. The mean profile based on 2015 spectra is plotted with a 
dashed line since the resolution in those spectra was lower. It also has the 
strongest emission contribution. The deepest blue wing absorption is seen in 
the mean profile from 2020 (dark green). However, also the 2022 profile 
(salmon) and the mean of the 2023 profiles (dark yellow) have a deep 
absorption in the blue wings.
}
\label{afig:Hbetamed}
\end{figure}

Similar to Fig.~\ref{afig:Hbetayr}, in Fig.~\ref{afig:Hbetamed}, we present 
individual and mean (for details, see Table~\ref{tab:vari}) line profiles of 
the H$\beta$ line based on medium-resolution observations. The observations 
were obtained with the Cassegrain Echelle Spectrograph (CES) at the Special 
Astrophysical Observatory (SAO) of the Russian Academy of Sciences in 2015. 
After 2015, the observations were carried out using the ShAO Fiber-Echelle 
Spectrograph (ShAFES), developed jointly by specialists from ShAO and SAO. 
For more details, we refer to
\citet{Hemayil}. 
From 2015, there are 22 spectra with a resolution $R\approx14\,000$. After 
2015, the resolution is higher ($R\approx28\,000$), but the number of spectra 
varies between one and seven per year. It has already been discussed by
\citet{Hemayil}
that the individual H$\beta$ line profiles obtained during 2015 show 
variability from night to night. The mean profile of 2015 (light blue) has the 
strongest emission component and the weakest absorption wings on both sides. 
The 2016 profile (brown) shows the weakest emission component whereas the 
absorption wings show an average behaviour.

The profile of 2020 (dark green) shows the strongest absorption in the 
blue wing. The single 2022 profile (salmon) and the average of 2023 (light 
yellow) also have both very deep absorptions in the blue wing. Also the 
emission line components are among the strongest, but the shapes are very 
different. In the 2022 profile the red emission peak is much higher than the 
blue one, whereas in the 2023 profile the difference in height is not that 
large, with the redder peak still being stronger.

\section{Variability of metal lines}\label{bsec:metvari}

\begin{figure}
\centering
\includegraphics[width=0.24\textwidth]{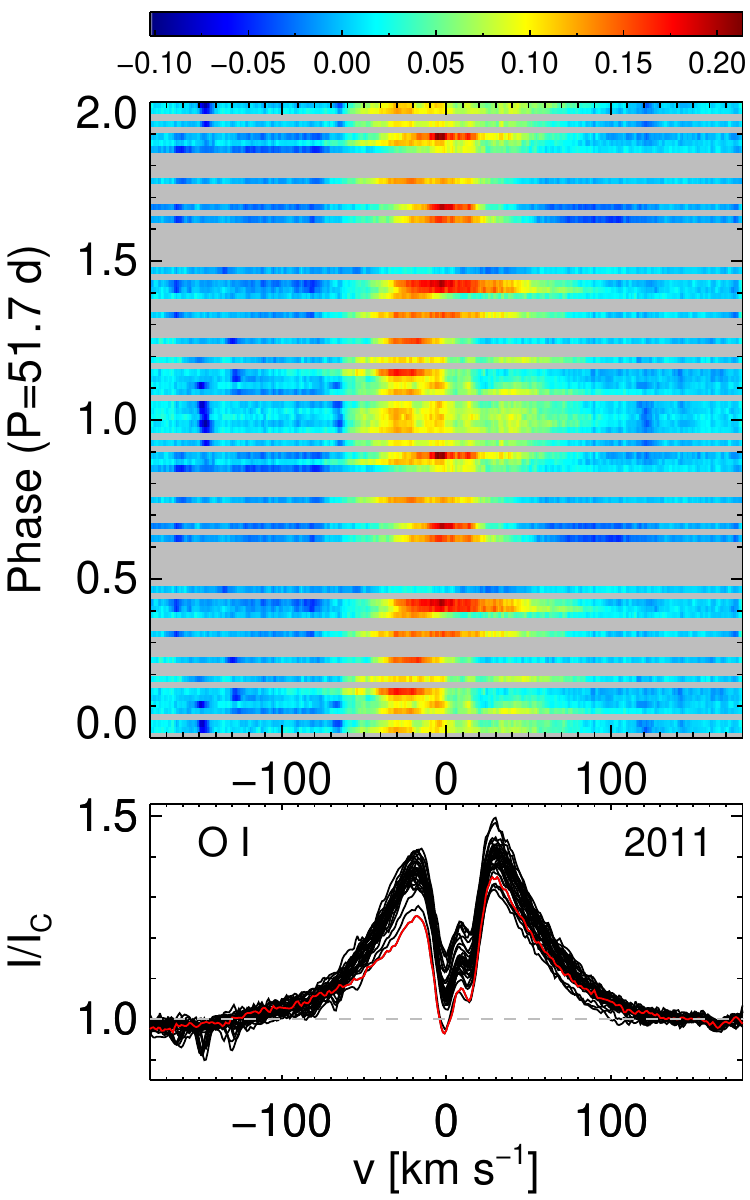}
\includegraphics[width=0.24\textwidth]{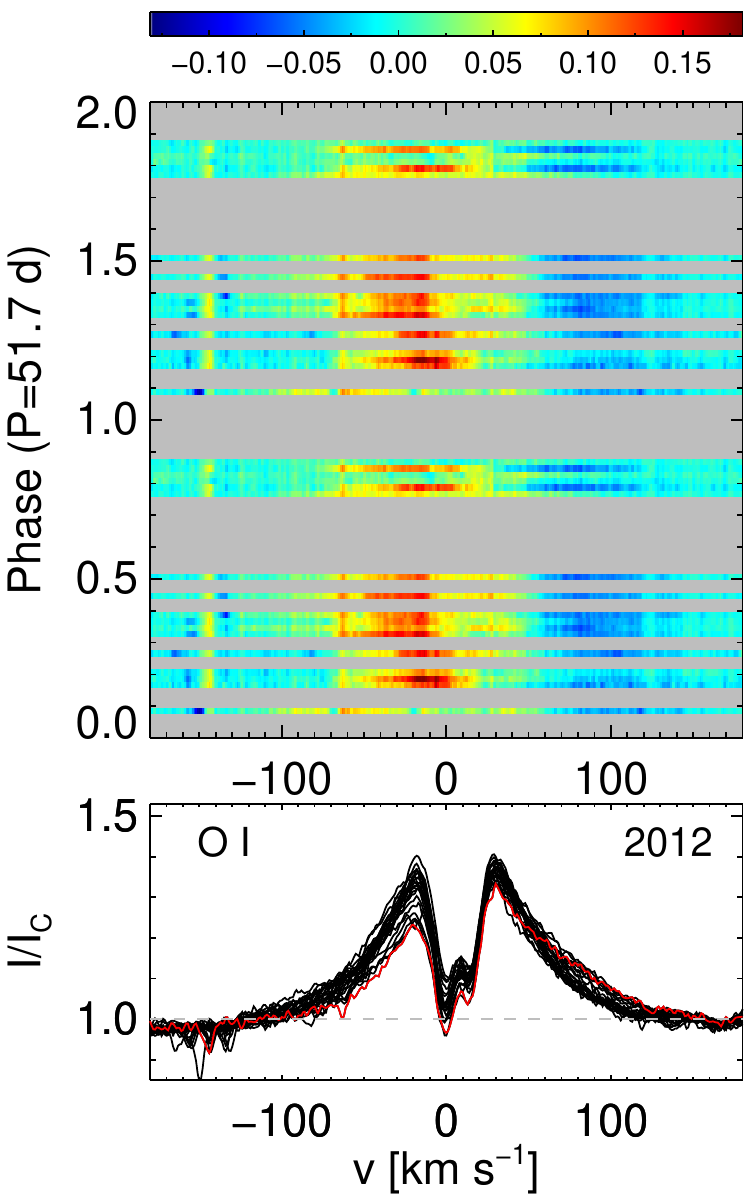}
\caption{
As Fig.~\ref{fig:Hdyndiff12}, but for the \ion{O}{I} 8446\,\AA{} line 
profiles. Dynamical spectra for the years 2011 (left) and 2012 (right) are 
constructed using differences between individual \ion{O}{I} 8446\,\AA{} line 
profiles and the line profile with the lowest core emission (shown in red). 
}
\label{afig:Odyndiff12}
\end{figure}

The metal lines of HD\,190073 are strongly contaminated by emission and show
also variability over time. The dynamical spectra and overplotted Narval 
profiles for 2011 and 2012 of the \ion{O}{i} line at 8447\,\AA{} are shown in
Fig.~\ref{afig:Odyndiff12}. The variability seen in \ion{O}{i} is not as 
clear as in the hydrogen lines, but still the 2011 data set shows that the 
emission strength is lower at phase 0 at the best visibility of the positive 
magnetic pole.

\begin{figure}
\centering
\includegraphics[width=0.32\textwidth]{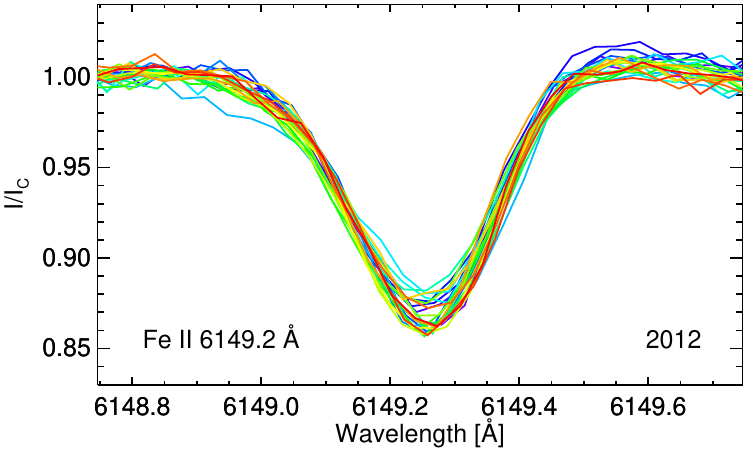}
\includegraphics[width=0.32\textwidth]{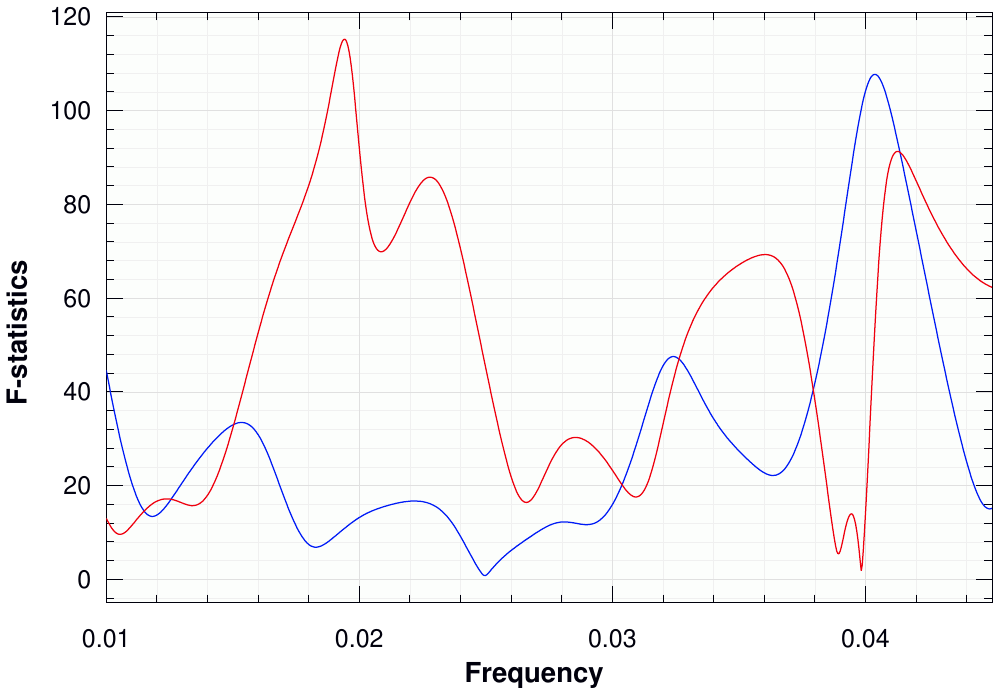}
\includegraphics[width=0.32\textwidth]{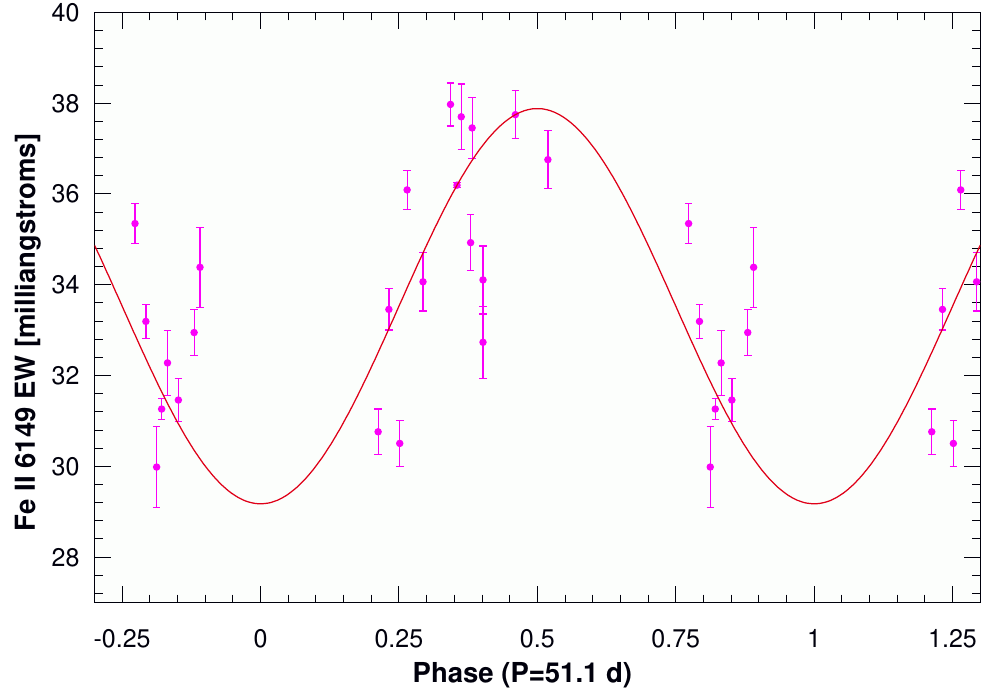}
\caption{
Variability of the \ion{Fe}{ii} 6149\,\AA{} lines. 
{\it Top:} 2012 Narval line profiles overplotted for \ion{Fe}{ii} 6149\,\AA{}.
{\it Middle:} Periodogram based on EWs measured from the profiles shown in
the top panel. The red line is for the observations and the blue line for
the window function.
{\it Bottom:} Magenta circles represent EWs measured from the 2012 Narval
spectra using the \ion{Fe}{ii} 6149\,\AA{} line and are phased with
$P=51.1$\,d. The obtained period is slightly shorter than the one obtained
from the mean longitudinal magnetic field measurements, but within the errors
still similar. The sinusoidal fit is shown with the red line.
}
\label{afig:Feoverplots}
\end{figure}

In the case of the metal absorption lines, the scale of variations is even 
smaller, but the equivalent widths of \ion{Fe}{ii} 6149\,\AA{} vary similar
to the \ion{O}{i} line. The line and EW variability of the \ion{Fe}{ii} line in
2011 Narval spectra is presented in Fig.~\ref{afig:Feoverplots}. The EWs 
were also used for a period search, which led to $P_{\rm rot}=51.1\pm0.6$\,d. 
This period is slightly shorter than the one obtained from the mean 
longitudinal magnetic field measurements $P_{\rm rot}=51.7\pm0.06$\,d, but 
is comparable within the errors.

\section{Dynamical spectrum using all 2011 Narval observations}

\begin{figure}
\centering
\includegraphics[width=0.26\textwidth]{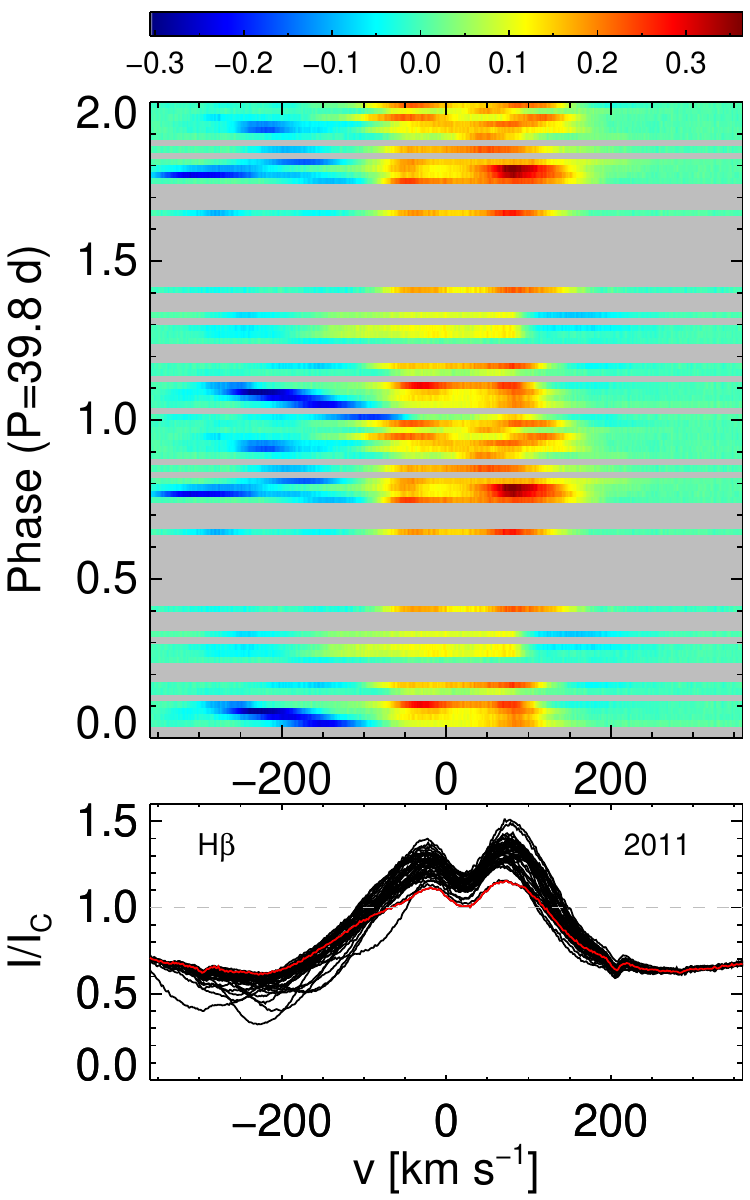}
\caption{
As Fig.~\ref{fig:Hdyndiff12}, but using Narval observations obtained in 2011
and the period reported by
\citet{Alecian2013b}. 
}
\label{afig:Hbeta39}
\end{figure}

In Fig.~\ref{afig:Hbeta39} we show the dynamical spectrum for the H$\beta$ line 
based on Narval observations obtained in 2011 and phased with the period
reported by
\citet{Alecian2013b}.

\end{appendix}

%

\end{document}